\documentclass[prd,twocolumn,showpacs,superscriptaddress,preprintnumbers,nofootinbib, amsmath,amssymb]{revtex4-2}
\usepackage{graphicx}
\usepackage{aas_macros}
\usepackage{orcidlink}

\begin{document}

\title{Strong-lensing Perturber Signatures in Self-interacting Dark Matter Simulations}

\author{Demao Kong\orcidlink{0000-0003-1723-8691}}
\email{dkong012@ucr.edu}
\affiliation{Department of Physics \& Astronomy, University of California, Riverside, CA 92521, USA}

\author{Ethan O.~Nadler\orcidlink{0000-0002-1182-3825}}
\email{enadler@ucsd.edu}
\affiliation{Department of Astronomy \& Astrophysics, University of California, San Diego, La Jolla, CA 92093, USA}

\author{Hai-Bo Yu\orcidlink{0000-0002-8421-8597}}
\email{haiboyu@ucr.edu}
\affiliation{Department of Physics \& Astronomy, University of California, Riverside, CA 92521, USA}

\date{\today}

\begin{abstract}
Motivated by recent detections of low-mass perturbers in strong gravitational lensing systems, we investigate analogs of these objects in the Concerto suite, a set of cosmological $N$-body zoom-in simulations of self-interacting dark matter (SIDM) with high-amplitude, velocity-dependent cross sections. We investigate characteristic halo properties relevant to gravitational imaging measurements, focusing on the projected enclosed mass and the central density slope. In SIDM, these quantities evolve continuously through gravothermal processes, spanning core-expansion and core-collapse phases, in sharp contrast to cold dark matter, where they remain nearly static after halo formation. This SIDM evolution further depends on tidal environment and merger history, which can be probed through strong lensing. We also identify simulated SIDM halos whose properties are consistent with the properties of low-mass perturbers inferred from recent observations, and we demonstrate that the core-collapse mechanism offers a compelling explanation for their observed high densities. Our results highlight the potential of strong gravitational lensing as a powerful probe of dark matter self-interactions.

\end{abstract}

\maketitle

\section{Introduction}
\label{sec:intro}

One of the important predictions of the cold dark matter (CDM) framework is that structure forms hierarchically: smaller structures merge into larger structures and become substructures~\cite{1993MNRAS.264..201K,1999ApJ...522...82K, Moore:1999nt}. Subhalos within a larger dark matter halo therefore provide a crucial testbed for probing the nature of dark matter, as its particle properties can significantly and distinctly impact predictions for subhalo density profiles and abundances; well-studied examples include warm dark matter~\cite{Bode:2000gq,Lovell:2013ola,Gilman:2019nap,Keeley:2024brx,Nadler:2025fcv} and self-interacting dark matter (SIDM)~\cite{Vogelsberger:2012ku,Rocha:2012jg,Gilman:2022ida,Fischer:2023lvl,Nadler:2023nrd,Despali:2025koj,Nadler:2024kyz,Nadler:2025jwh,Hou:2025gmv}. The observation or non-observation of subhalos with certain properties can provide constraints on different dark matter models. Thus, the ability to detect substructure is critical.

Gravitational lensing provides a channel for subhalo detection that does not depend on the presence of baryonic matter. In particular, perturbations in magnified images can probe the mass of perturbers near strong lenses \cite{Mao:1997ek, Metcalf:2001ap, Koopmans:2005nr}. In theory, such detections depend on the mass of the perturber enclosed within a specific region, and can therefore probe subhalos at the low-mass end that may not host galaxies~\cite{Benitez-Llambay:2020zbo,Nadler:2025mnz}. There have been many studies trying to detect subhalos in different lensing systems~\cite{Dalal:2001fq, Bradac:2001mv, Metcalf:2001es, 2010MNRAS.408.1969V, Vegetti:2012mc, Oguri:2012vg, MacLeod:2012sv, Nierenberg:2014cga, Hezaveh:2016ltk, Minor:2020hic,Sengul:2022edu,Lange:2024pef, Enzi:2024ygw, Despali:2024ihn, Tajalli:2025qjx, Li:2025kpb, Cao:2025eff, He:2025wco, Ephremidze:2025mqg}; see~\cite{Vegetti:2023mgp} for a review. By comparing the halo properties predicted by simulations to the properties of observed perturbers, we can therefore gain insights into alternative dark matter models. For example, the subhalo perturber of the strong lens SDSS J0946\textup{+}1006 (J0946)~\cite{2010MNRAS.408.1969V} has a surprisingly high density compared to CDM predictions~\cite{Minor:2020hic,Nightingale:2022bhh,Ballard:2023fgi,Enzi:2024ygw,Despali:2024ihn,Minor:2025,Tajalli:2025qjx}, motivating explanations based on SIDM models~\cite{Nadler:2023nrd,Tajalli:2025qjx,Li:2025kpb}.

SIDM introduces self-interactions between dark matter particles, driving halos to undergo a two-stage gravothermal evolution: core expansion, in which the collisional thermalization forms a shallow isothermal core, and core collapse, where the central density rises to match or exceed that of a CDM halo; see~\cite{Tulin:2017ara,Adhikari:2022sbh} for reviews and reference therein. Core-collapsed SIDM halos can be natural candidates for high-concentration lensing perturbers that deviate from CDM predictions, making their lensing properties a sensitive probe of the SIDM parameter space. Ref.~\cite{Nadler:2023nrd} performed high-resolution cosmological simulations of a strong lens-scale system and indeed found that the core-collapse mechanism can account for the high density of the J0946 perturber.

In this SIDM scenario, both the cross-section amplitude ($\sigma_0/m$) and turnover velocity ($w$) of Yukawa SIDM models~\cite{Feng:2009hw,Tulin:2017ara} can be constrained. Ref.~\cite{Nadler:2023nrd} assumed that $\sigma_{0}/m = 147.1~\rm{cm^2/g}$ and $w=120~{\rm cm^2/g}$. Ref.~\cite{Yang:2024uqb} further used a parametric method~\cite{Yang:2023jwn} to extend the original study by exploring a wider SIDM model parameter space. Together, Refs.~\cite{Nadler:2023nrd,Yang:2024uqb} found that for $\sigma_0/m\sim70\textup{--}100~{\rm cm^2/g}$ and $w\sim100~{\rm km/s}$, strong lens subhalos with masses $\sim10^{10}~{\rm M_\odot}$, can be deeply core-collapsed, resulting in much more compact, dense inner structures than their CDM counterparts, broadly consistent with the inferred properties of the J0946 perturber. Moreover, for velocity-dependent SIDM models, the fraction of core-collapsed halos peaks at a characteristic mass scale determined by the SIDM cross section~\cite{Ando:2024kpk,Nadler:2025jwh,Shah:2023qcw}, which can be used to further discern SIDM models using observations over a wide mass range. In addition, the SIDM models favored by lensing observations also predict sharply rising rotation curves in dark-matter-dominated spiral galaxies~\cite{Roberts:2024uyw,Kong:2025irr}, as well as high densities in small dark subhalos of mass $\sim10^6\textup{--}10^{8}~{\rm M_\odot}$ capable of perturbing stellar streams~\cite{Zhang:2024fib,Fischer:2025rky}.

In this work, we present a comprehensive investigation of simulated analogs for all known strong lensing perturber candidates detected through gravitational imaging. Besides J0946, we also include possible perturber detections in three more gravitational lensing systems: JVAS B1938\textup{+}666 (B1938)~\cite{Vegetti:2012mc,Sengul:2021lxe, Sengul:2022edu, Despali:2024ihn,Tajalli:2025qjx}, SDP.81~\cite{Hezaveh:2016ltk}\footnote{A recent study~\cite{Stacey:2025} did not confirm the detection of a subhalo perturber in the SDP.81 system. In this work, we still include it as a reference case and will remind the reader of the conclusions from~\cite{Stacey:2025} whenever relevant.}, and SPT2147-50~\cite{Lange:2024pef}. To compare the inferred properties of these perturbers against theoretical predictions, we explore the lensing characteristics of halos from the zoom-in SIDM Concerto $N$-body simulation suite~\cite{Nadler:2025jwh}, which incorporates three velocity-dependent SIDM models and covers a range of main host halo masses $\sim10^{11}\textup{--}10^{14}~{\rm M_\odot}$. We show that SIDM halos' projected enclosed masses and logarithmic density slopes evolve over the course of gravothermal evolution and compare them with those from CDM simulations. In particular, the evolution path for core-collapsed halos can be characterized in two stages, which are connected by a turnover point that maximizes enclosed mass within a characteristic inner radius probed by strong lensing data. We also adopt the perturber models from recent studies as reference points to search for possible analogs in the simulations, showing that core-collapsed SIDM halos can provide a reasonable explanation for all observed strong lensing perturbers.

The rest of the paper is organized as follows. In Sec.~\ref{sec:method}, we describe the SIDM Concerto zoom-in simulations, along with the methods used to calculate projected enclosed masses and density slopes; we also explore environmental effects of host systems. In Sec.~\ref{sec:subevo}, we examine the evolution of subhalo masses and density slopes, including case studies that highlight the impact of mergers and tidal stripping. We show that tidal forces can accelerate gravothermal evolution relative to field halos, while mergers can delay core collapse in field environments. In Sec.~\ref{sec:subana}, we identify simulated analogs of observed lensing perturbers and discuss modeling uncertainties. Finally, in Sec.~\ref{sec:discon}, we present our discussion and conclusions. Appendix~\ref{appex:cv} provides results based on an alternative concentration definition. { Appendix~\ref{appex: medrho} presents the median 3D density profiles and corresponding $1\sigma$ scatter for simulated subhalos.
}

\section{Data and Methods}
\label{sec:method}

\subsection{Simulations}
\label{subsec:sim}

\begin{figure}[!]
    \centering
    \includegraphics[width=\linewidth]{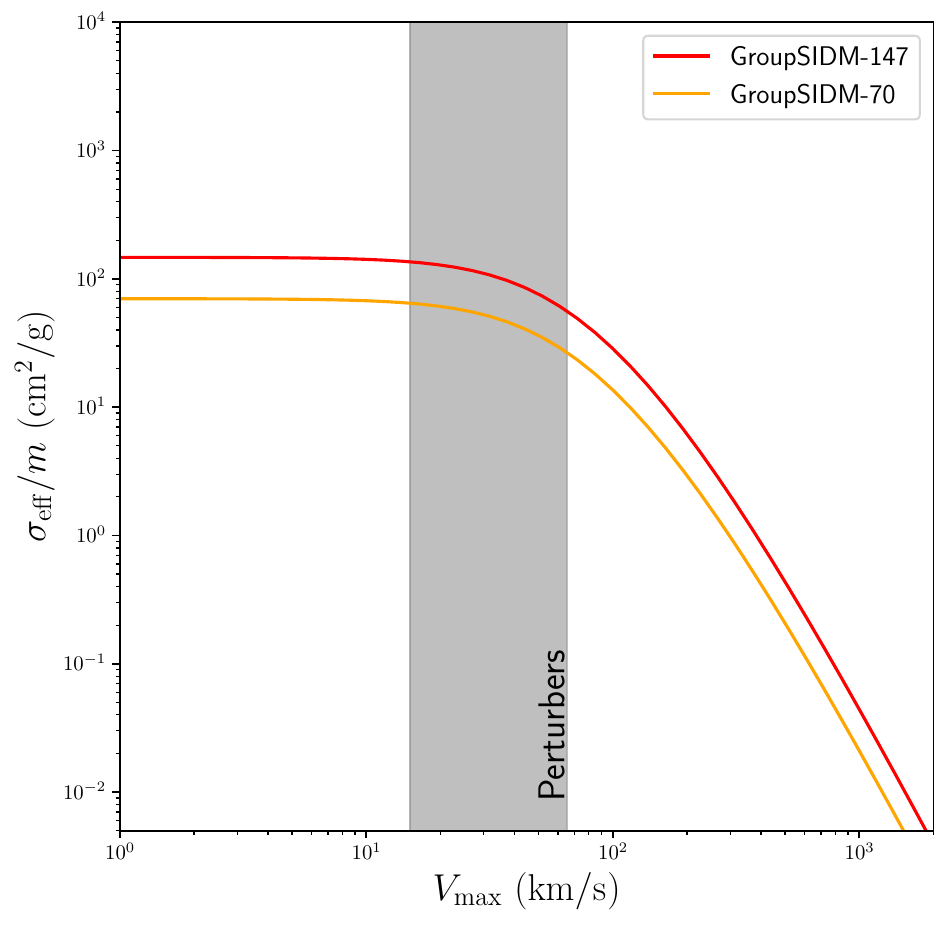}
    \caption{Effective dark matter self-interaction cross sections as a function of the halo maximum circular velocity in GroupSIDM-70 (orange) and GroupSIDM-147 (red) models. The shaded region shows the $V_{\rm max}$ range derived from the observed strong-lensing perturber models we consider in this work.}
    \label{fig:sigma-vmax}
\end{figure}

For this work, we mainly focus on the Group strong lens analog Halo352 from the zoom-in SIDM Concerto simulation suite~\cite{Nadler:2025jwh}, since its host mass is $M_{\rm host}=1.3\times10^{13}~\rm{M_{\odot}}$, which is typical for a strong-lensing system~\cite{Gavazzi:2007vw,Auger:2010va}. The Group suite contains three different dark matter models: CDM, GroupSIDM-70, and GroupSIDM-147, as described below. The self-interactions are modeled with a differential scattering cross section~\cite{Ibe:2009mk, Yang:2022hkm}: 
\begin{equation}
\frac{\mathrm{d} \sigma}{\mathrm{~d} \cos \theta}=\frac{\sigma_0 w^4}{2\left[w^2+v^2 \sin ^2(\theta / 2)\right]^2},
\label{eq:yukawa}
\end{equation}
where $v$ and $\theta$ denote the relative velocity and scattering angle, respectively, $\sigma_{0}$ is the cross-section amplitude, and $w$ is the turnover velocity characterizing the transition from $\sigma \propto v^{4}$ to $\sigma \propto v^{0}$. The GroupSIDM-70 model assumes $\sigma_{0}/m = 70~\rm{cm^2/g}$ and the GroupSIDM-147 model assumes $\sigma_{0}/m = 147.1~\rm{cm^2/g}$, with both adopting $w=120~\rm{km/s}$. Fig.~\ref{fig:sigma-vmax} shows the effective cross section~\cite{Yang:2022hkm,Yang:2022mxl,Nadler:2023nrd} as a function of the halo maximum circular velocity $V_{\rm max}$ for the GroupSIDM-70 (red) and GroupSIDM-147 (orange) models, along with the $V_{\rm max}$ range relevant for strong lensing perturbers (shaded gray). The SIDM simulations employ the viscosity cross section, defined by weighting the differential cross section in Eq.~\ref{eq:yukawa} with $\sin^2\theta$; see~\cite{Yang:2022hkm} for details.

The Group suite has a simulation particle mass of $4\times10^{5}~\rm{M_{\odot}}$ and a Plummer-equivalent softening length of $\epsilon=0.24~\rm{kpc}$. We analyze subhalos of the host halo as well as field halos within a distance of $6~\mathrm{Mpc}$, where they are resolved with high-resolution particles. Bound particles of each halo are identified through an iterative unbinding procedure, requiring particles to have negative total energy (kinetic plus potential). For subhalo analyses, we combine catalogs from the Rockstar-plus-Consistent-Trees~\citep{Behroozi:2011ju,Behroozi:2011js} and Symfind~\citep{Mansfield:2023prs} halo finders to ensure completeness, following the recommendation of~\cite{Kong:2025kkt}. Virial masses $M_{\rm vir}$ are defined according to the Bryan–Norman criterion~\cite{Bryan:1997dn}; for subhalos, only bound particles are included in the mass measurement.

\begin{table}
\caption{\label{tab:simspec}SIDM Concerto Simulation Parameters}
\begin{ruledtabular}
\begin{tabular}{l|l|l|l}
Halo & $m_{p}$~[$\rm{M_{\odot}}$] & $\epsilon$~[\rm{kpc}] & SIDM Model \\
\hline
Halo352 (Group) & $4.0\times10^{5}$ & $0.24$ & GroupSIDM-70/147 \\
Halo004 (MW) & $5.0\times10^{4}$ & $0.11$ & GroupSIDM-147 \\
Halo104 (LMC) & $6.3\times10^{3}$ & $0.06$ & GroupSIDM-147 \\
\end{tabular}
\end{ruledtabular}
\end{table}

For the perturber detection associated with B1938, the Group simulations lack the mass and spatial resolution necessary to resolve the inner regions of halos that contribute to the observed lensing signal. To address this, we also analyze SIDM Concerto simulations of a Milky Way analog (MW; Halo004, with the CDM counterpart originally presented in~\cite{Buch:2024ssx}) and a Large Magellanic Cloud analog (LMC; Halo104, with the CDM counterpart in~\cite{Nadler:2022dvo}); their simulation parameters are summarized in Table~\ref{tab:simspec}. When analyzing field halos around these hosts, we include systems within $3~\mathrm{Mpc}$ of the MW center and $1.5~\mathrm{Mpc}$ of the LMC center, ensuring that only high-resolution particles are used and contamination from low-resolution regions is avoided.

While the Group host resides in a large-scale overdense region of the cosmic environment, the MW and LMC hosts are located in underdense regions~\cite{Nadler:2022dvo,Nadler:2025mnz}. Thus, it is important to examine how different environments affect the structural properties of halos in these simulations. We calculate the effective concentration of simulated halos~\cite{Yang:2022mxl}, defined as
\begin{equation}
c_{\rm eff}=\frac{r_{\rm vir}}{r_{\rm max}/2.1626}
\label{eq:ceff}
\end{equation}
where $r_{\rm vir}$ is the virial radius and $r_{\rm max}$ is the radius at which the maximum circular velocity occurs. In the limit of a Navarro–Frenk–White (NFW) profile~\cite{Navarro:1995iw}, $r_{\rm max}=2.1626r_s$, and the familiar definition of halo concentration, $c_{\rm vir}=r_{\rm vir}/r_s$, is recovered. For SIDM halos in the core-expansion phase, $r_{\rm max}$ can increase mildly, leading to slightly smaller $c_{\rm eff}$ values than their CDM counterparts. By contrast, for core-collapsed SIDM halos, $r_{\rm max}$ can be significantly reduced, yielding much larger $c_{\rm eff}$ values than in CDM. Hence, the effective concentration is particularly useful for characterizing core-collapsed halos~\cite{Yang:2022mxl} that are the main focus of this work.

\begin{figure}[!]
    \centering
    \includegraphics[scale=0.48]{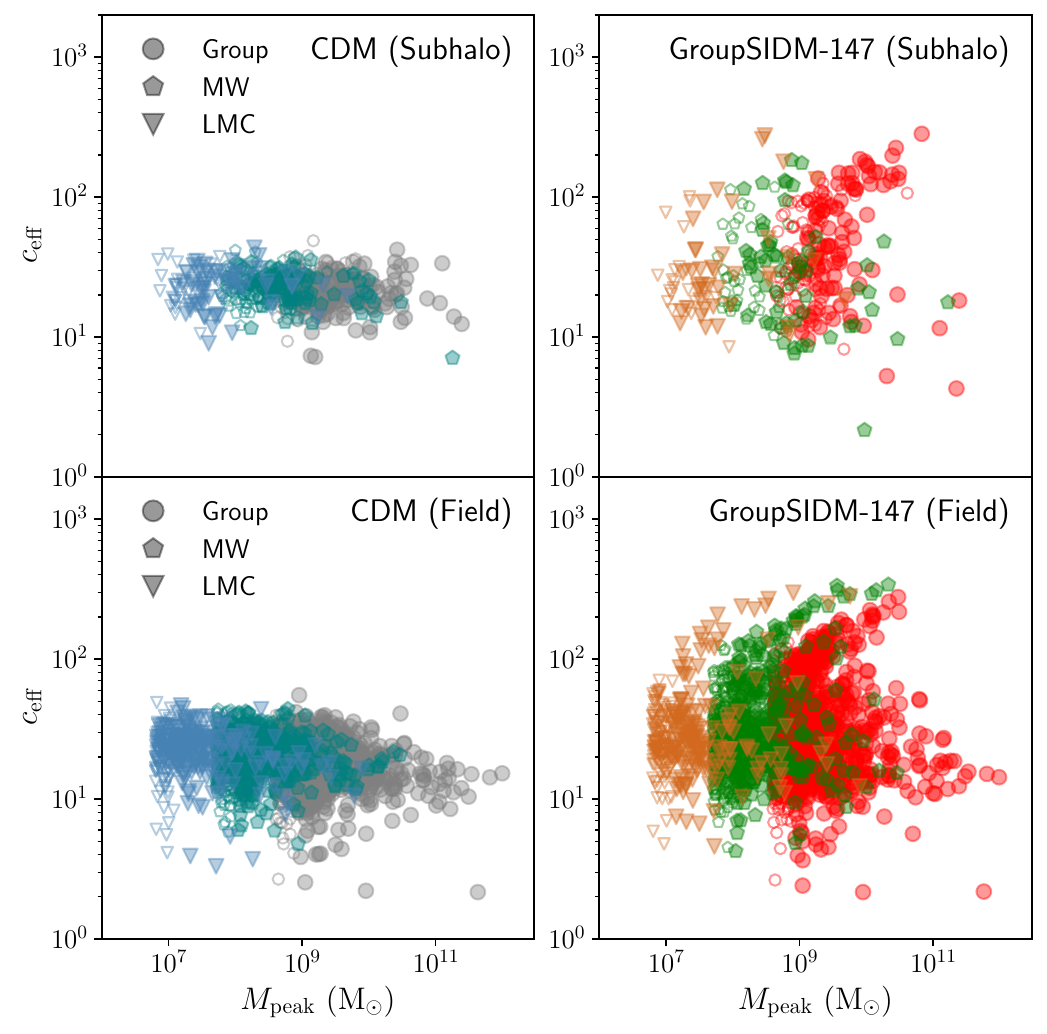}
    \caption{Distributions of effective concentration (Eq.~\ref{eq:ceff}) as a function of peak mass for subhalos (top) and field halos (bottom) in the Group (circles), MW (pentagons), and LMC (triangles) simulations, shown for CDM (left) and GroupSIDM-147 (right). Halos with more than $2000$ particles are shown as large filled markers, and those with $1000\textup{--}2000$ particles as small open markers.}
    \label{fig:ceff}
\end{figure}

In Fig.~\ref{fig:ceff}, we show the distributions of the effective concentration of subhalos (top) and field halos (bottom) with respect to their peak masses in our Group (circles), MW (pentagons), and LMC (triangles) systems, for CDM (left) and GroupSIDM-147 (right). The peak mass $M_{\rm peak}$ is defined as the maximum mass a halo attains over the course of the simulation. For subhalos, $M_{\rm peak}$ typically occurs before they fall into the host system, whereas for field halos it usually occurs at $z=0$. We clearly see that SIDM halos exhibit much larger scatter in $c_{\rm eff}$ at fixed $M_{\rm peak}$ compared to CDM. In particular, core-collapsed halos can reach very high $c_{\rm eff}$ values, often exceeding $100$. Moreover, the $c_{\rm eff}$--$M_{\rm peak}$ relation extends smoothly across the four host masses, indicating that biases from the differing environments of Concerto hosts are minimal. For this reason, we will identify and present analogs of the B1938 perturber from the LMC system, which has the highest resolution. We have also verified that the overall results remain unchanged when using the MW system. { In Appendix~\ref{appex:cv}, we repeat the analysis using an alternative definition of halo concentration and find that our conclusions remain unchanged. }

\begin{figure*}[t]
    \centering
    \includegraphics[width=\linewidth]{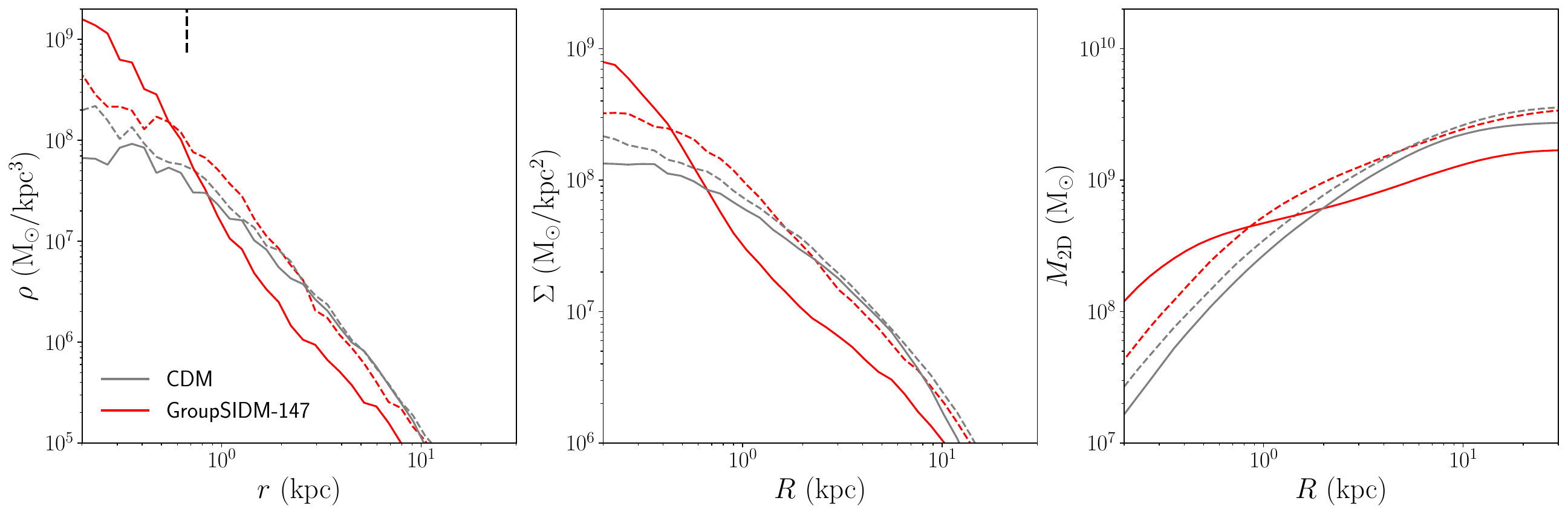}
    \caption{Profiles of the 3D density (left), surface density (middle) and projected enclosed mass (right) for two representative SIDM subhalos (red) and their CDM counterparts (gray) at $z=0$ in Group CDM and GroupSIDM-147 simulations. The solid and dashed curves denote subhalos with masses $2\times10^{9}~{\rm M_\odot}$ and $3.5\times10^{9}~{\rm M_\odot}$ at $z=0$, respectively. In the left panel, the vertical dashed line marks the resolution limit $2.8\epsilon = 0.67~\mathrm{kpc}$.}
    \label{fig:example}
\end{figure*}

\subsection{Projected enclosed mass and density slope}
\label{subsec:gammamethod}

Strong lensing measurements can constrain both the projected enclosed mass $M_{\rm 2D}$ and the projected logarithmic density slope $\gamma_{\rm 2D}$ at a characteristic radius. For J0946, this radius is $R \approx 1~{\rm kpc}$~\cite{Minor:2020hic, Despali:2024ihn}. Motivated by this, we evaluate $M_{\rm 2D}$ and $\gamma_{\rm 2D}$ for subhalos in the Group simulations at $R=1~\mathrm{kpc}$. For each subhalo, we generate random lines of sight by rotating particle positions according to Euler angles $\theta$, $\phi$, and $\psi$. Taking the line of sight as the $z$-axis and assuming an axis ratio $q=1$ in the $x$–$y$ plane, the projected enclosed mass at radius $R$ is obtained by sorting particles according to their projected positions, $R = \sqrt{x^{2}+y^{2}}$. The surface density profile $\Sigma(R)$ is then measured by binning particles in logarithmically spaced radial bins. Finally, the projected logarithmic density slope is defined as $\gamma_{2\mathrm{D}}\equiv \mathrm{d}\log \Sigma(R)/\mathrm{d}\log R$, which we evaluate by averaging over the radial range $R=1\pm0.25~\rm{kpc}$.

We impose a minimum resolution requirement of $1000$ particles per subhalo, corresponding to a mass cut of $M > 4\times10^{8}~\rm{M_{\odot}}$ in the Group simulations. For each selected subhalo, we generate $1000$ randomized lines of sight and average the results to suppress numerical noise. The characteristic radius $R=1~{\rm kpc}$ motivated by the J0946 perturber is safely above the resolution limit of $2.8\epsilon=0.67~{\rm kpc}$ set by the force softening length. For other strong lensing systems considered in this work, we directly compare the inferred density profiles of the perturbers with those derived from our simulations.

\section{Density Profile Measurements}
\label{sec:subevo}

\subsection{Representative Examples}
\label{subsec: rep}

In Fig.~\ref{fig:example}, we show two example subhalos from the Group simulations in the GroupSIDM-147 and CDM models, comparing their 3D density, surface density, and projected enclosed mass profiles (left to right panels). The halo masses are $2\times10^{9}~{\rm M_\odot}$ (solid) and $3.5\times10^{9}~{\rm M_\odot}$ (dashed) at $z=0$. The $2\times10^{9}~{\rm M_\odot}$ SIDM subhalo is deeply core-collapsed, with its central density reaching $10^9~{\rm M_\odot/kpc^{3}}$, roughly an order of magnitude higher than its CDM counterpart. At larger radii ($r>1~{\rm kpc}$), the SIDM subhalo shows reduced density relative to CDM, as mass is funneled inward during core collapse. The surface density and projected enclosed mass profiles display the same trend: the SIDM subhalo is more concentrated, has a steeper slope, and a higher enclosed mass at $R=1~{\rm kpc}$ compared to CDM. The $3.5\times10^{9}~{\rm M_\odot}$ SIDM subhalo (dashed) undergoes only mild core collapse, so the differences from its CDM counterpart are smaller, but the overall trend remains.

These results highlight that core-collapsed SIDM subhalos are more likely than their CDM counterparts to produce significant perturbations in strong lensing systems, and their inner structure near the characteristic radius can be directly probed by lensing measurements. In contrast to CDM subhalos, which remain nearly static after formation aside from tidal stripping, SIDM subhalos continuously evolve through different gravothermal phases. Consequently, we expect the characteristic lensing observables, $\gamma_{\rm 2D}$ and $M_{\rm 2D}$, to evolve over time in a manner distinct from CDM, as we demonstrate in the following section.

{ In Appendix~\ref{appex: medrho}, we show the median 3D density profiles and corresponding $1\sigma$ scatter at $z=0$ for all subhalos with masses $10^{9}\textup{--}10^{10}~\mathrm{M_\odot}$ in our Group simulations. Relative to CDM, SIDM subhalos exhibit a much broader distribution of inner densities due to gravothermal evolution, which amplifies the intrinsic halo-to-halo scatter and increases the central $1\sigma$ dispersion relevant for strong-lensing constraints by nearly an order of magnitude. While the median central density in GroupSIDM-70 remains similar to that in CDM, the median density in GroupSIDM-147 is higher, reflecting the more advanced gravothermal collapse of its subhalo population.
}

\subsection{Density slopes at different redshifts}
\label{subsec:gamma}

\begin{figure*}[t]
    \centering
    \includegraphics[width=\linewidth]{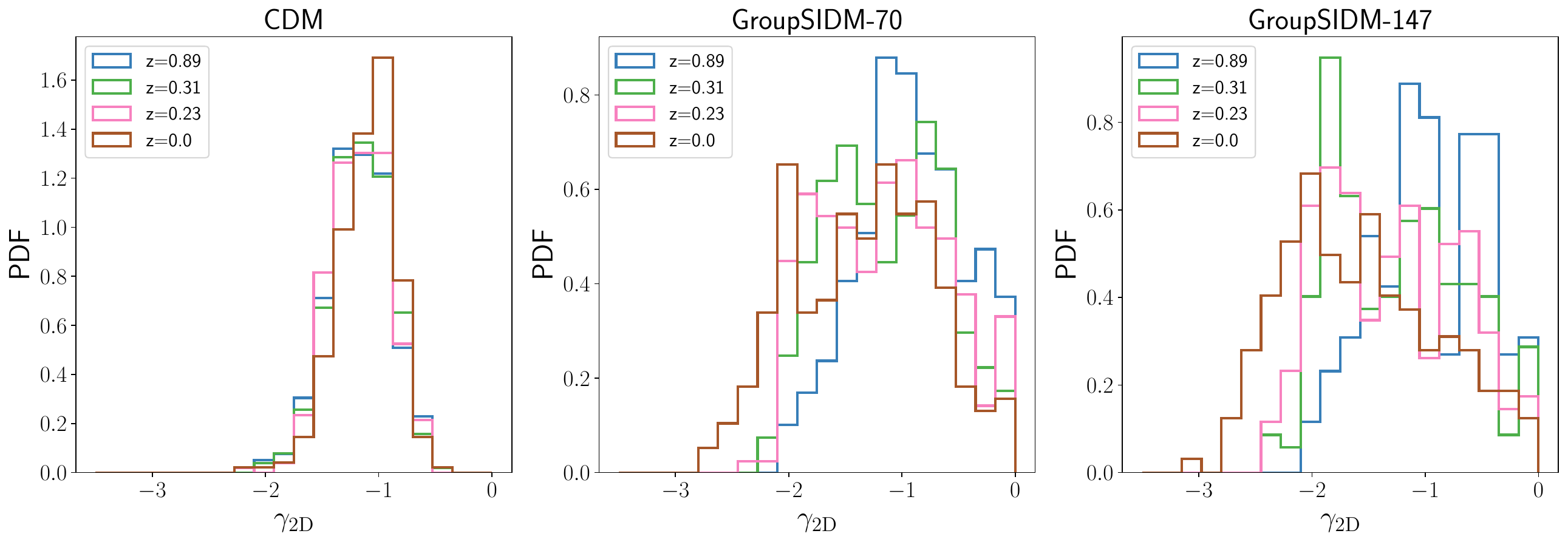}
    \caption{Probability distributions of the projected density slope $\gamma_{\rm 2D}$ of subhalos at redshifts $z=0.89$ (blue), $0.31$ (green), $0.23$ (pink), and $0$ (brown) in the Group simulations for CDM (left), GroupSIDM-70 (middle), and GroupSIDM-147 (right). We include all subhalos resolved with more than $1000$ particles, corresponding to masses above $4\times10^{8}~{\rm M_\odot}$. }
    \label{fig:sub-evo}
\end{figure*}

We now explore the evolution of the lensing characteristics of subhalos in the Group simulations by presenting distributions of $\gamma_{\rm 2D}$ at different redshifts. We choose simulation snapshots closest to the following possible strong lensing perturber detections: $z=0.89$ motivated by B1938 ($z_{\rm lens}=0.881$~\cite{Tonry:1999pp, Riechers:2011tz}), $z=0.31$ by {SDP.81} ($z_{\rm lens}=0.3$~\cite{2015PASJ...67...72T, Hezaveh:2016ltk}), and $z=0.23$ by J0946 ($z_{\rm lens}=0.222$ \cite{Gavazzi:2008aq}). Note the snapshot of $z=0.89$ is only $0.2~\rm{Gyr}$ away from the lens redshift of SPT2147-50 ($z_{\rm lens}=0.845$ \cite{Lange:2024pef}). We also include results from the $z=0$ simulation snapshot for reference. We evaluate $\gamma_{\rm 2D}$ at the characteristic radius $R=1~{\rm kpc}$, motivated by the system J0946~\cite{Minor:2020hic, Despali:2024ihn}.

In Fig.~\ref{fig:sub-evo}, we present probability distributions of the subhalo projected logarithmic density slope $\gamma_{\rm 2D}$ at redshifts close to $z_{\rm lens}$ for each lensing system of interest. In the left panel, we present the subhalos from the CDM simulation, showing that the distribution of density slopes peaks around $\gamma_{\rm 2D}\sim -1$ at different redshifts, and spans a relatively narrow range from $-2$ to $-0.5$ that exhibits little evolution with time. The origin of this range can be understood as follows. If the inner 3D density profile is approximated by a single power law, $\rho\propto r^{\gamma_{\rm 3D}}$, then $\gamma_{\rm 2D}\approx \gamma_{\rm 3D}+1$. The peak around $\gamma_{\rm 2D}\sim-1$ thus corresponds to $\gamma_{\rm 3D}\sim-2$, consistent with the intermediate region of NFW-like halos, where $\rho\propto r^{-2}$ and most subhalos have scale radii $r_s\sim 1~{\rm kpc}$. For subhalos with $r_s\ll 1~{\rm kpc}$, $\gamma_{\rm 2D}$ at $R=1~{\rm kpc}$ probes the outer region of NFW-like halos with $\rho\propto r^{-3}$, yielding values close to $-2$, while for those with $r_s>1~{\rm kpc}$, $\gamma_{\rm 2D}$ approaches $\sim-0.5$ .

In the middle panel of Fig.~\ref{fig:sub-evo}, we show that subhalos in the GroupSIDM-70 simulation have evolving density slopes. At $z=0.89$ where the simulation time is $t\sim 9~\rm{Gyr}$, the subhalos have a significant peak around $\gamma_{\rm 2D}\sim -1$ and a minor peak around $\gamma_{\rm 2D}\sim -0.3$, suggesting that the subhalo population is mainly composed of NFW-like systems, with a subdominant contribution from core-forming subhalos. As SIDM gravothermal evolution continues, the subhalos at $z=0.3$ ($t\sim 12~\rm{Gyr}$) become more cuspy and the peak moves from $\gamma_{\rm 2D}\sim -1$ to $\gamma_{\rm 2D}\sim -2$ due to core collapse. At $z=0$, the SIDM subhalos have a wide probability distribution spanning $\gamma_{\rm 2D}\sim -3$ to $\gamma_{\rm 2D}\sim 0$, reflecting their diverse density profiles.  

For comparison, in the right panel of Fig.~\ref{fig:sub-evo}, the subhalos in the GroupSIDM-147 simulation show a similar trend but with more rapid gravothermal evolution, consistent with the expectation due to the larger cross-section amplitude in this model. At $z=0.89$, the probability distribution has two peaks centered around $\gamma_{\rm 2D}\sim -0.5$ and $\gamma_{\rm 2D}\sim -1$. At $z=0.3$, many subhalos are deeply core-collapsed, leading to a narrow peak at $\gamma_{\rm 2D}\sim -2$. Some subhalos are still in the evolution process at this redshift, as suggested by a small peak around $\gamma_{\rm 2D}\sim -1$. At $z=0$, after a few more Gyr elapse, the overall population becomes more dense and the small peak around $\gamma_{\rm 2D}\sim -1$ disappears as the majority of subhalos become core-collapsed.

\subsection{Tidal Evolution Effects}
\label{subsec:tidal}

\begin{figure*}[t]
    \centering
    \includegraphics[width=\linewidth]{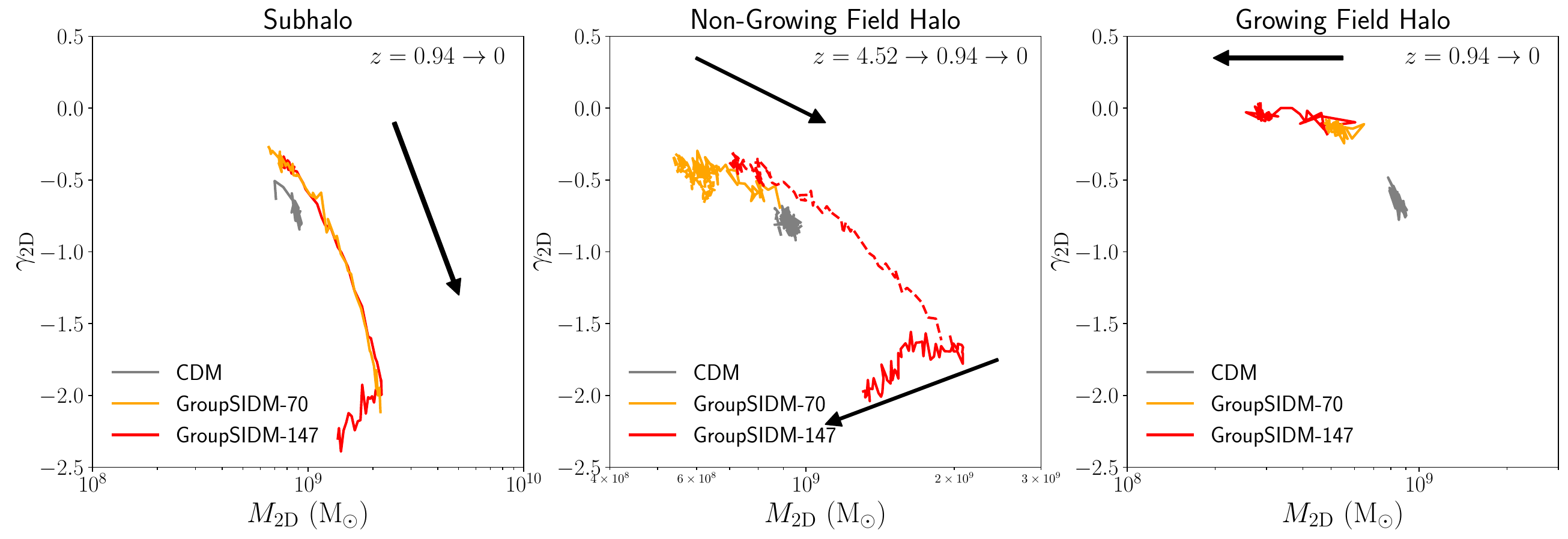}
    \caption{Evolution of the projected enclosed mass $M_{\rm 2D}$ and density slope $\gamma_{\rm 2D}$ and  for example subhalos (left), non-growing field halos (middle), and growing field halos (right). The subhalos and growing field halos evolve from $z=0.94$ to $z=0$ (solid). The non-growing halos evolve from $z=4.52$ to $z=0.94$ (dashed) and then to $z=0$ (solid). Gray, orange, and red lines correspond to CDM, GroupSIDM-70, and GroupSIDM-147 halos, respectively. Black arrows indicate the direction of evolution.}
    \label{fig:time-evo}
\end{figure*}

We further examine the time evolution of the projected enclosed mass and density slope through a case study. In the left panel of Fig.~\ref{fig:time-evo}, we focus on subhalos with $M_{\rm vir}\approx2.3\times10^{10}~\rm{M_{\odot}}$ at infall $z=0.94$. To enable a one-to-one comparison, we match subhalos across the three simulations based on their pre-infall trajectories and mass assembly histories. For the CDM case, the density slope exhibits only minor changes during tidal evolution, consistent with the results of Sec.~\ref{subsec:gamma}. By contrast, in the SIDM runs the evolutionary tracks of the GroupSIDM-70 and GroupSIDM-147 subhalos initially coincide, but the GroupSIDM-147 system evolves along this track more rapidly due to its larger cross section. Notably, the GroupSIDM-147 subhalo undergoes a turnover at $\gamma_{\rm 2D}\sim-2$, leading to a smaller enclosed mass but a steeper central density slope at later times.

To investigate inner profile evolution in the absence of tidal effects, we analyze field halos with quiescent merger histories but similar properties to the subhalos in the left panel at $z=0.94$. The middle panel of Fig.~\ref{fig:time-evo} presents halos with minimal mass growth from $z=0.94$ to $z=0$, shown as solid lines. To extend their evolutionary histories, we trace the merger trees back to $z=4.52$, with the interval between $z=4.52$ and $z=0.94$ shown as dashed lines. For the CDM halo, the enclosed mass and density slope remain essentially unchanged, as expected. For the GroupSIDM-70 halo, the profile undergoes core expansion between $z=4.52$ and $z=0.94$, and by the end of the simulation its enclosed mass and density slope are broadly comparable to those of the CDM halo as it approaches the onset of core collapse. Note that these field halos grow from $2.3\times10^{10}~{\rm M_\odot}$ at $z=0.94$ to $3\times10^{10}~{\rm M_\odot}$ at $z=0$.

Interestingly, the corresponding GroupSIDM-147 field halo exhibits a history similar to that of the GroupSIDM-147 subhalo shown in the left panel of Fig.~\ref{fig:time-evo}. In particular, it displays a two-stage evolution: initially, the enclosed mass increases along with the density slope, followed by a turnover at $\gamma_{\rm 2D}\approx-1.7$, after which the enclosed mass decreases while the density slope continues to rise. Although the qualitative behavior is similar in a non-tidal environment, the field halo traverses this evolutionary path much more slowly, over the full interval from $z=4.52$ to $z=0$, whereas the subhalo completes a similar path between $z=0.94$ and $z=0$. This comparison suggests that tidal effects can accelerate gravothermal evolution~\cite{Nishikawa:2019lsc,Sameie:2019zfo,Zeng:2021ldo} and drive the turnover points we identify, where subhalos develop steeper slopes while reaching comparable maximum enclosed masses relative to their field counterparts.

A recent study~\cite{Li:2025kpb} utilized the isothermal Jeans model~\cite{Kaplinghat:2015aga,Jiang:2022aqw} to explore the evolution of isolated SIDM halos, obtaining similar qualitative turnover behavior. Quantitatively, the turnover points in~\cite{Li:2025kpb} are at $\gamma_{\rm 2D}\sim-1.5$. This slope is overall consistent with our findings for the field GroupSIDM-147 halo shown in the middle panel of Fig.~\ref{fig:time-evo}, but less steep than the $\gamma_{\rm 2D}\sim -2$ turnover point for the GroupSIDM-147 subhalo in the left panel. The difference may arise because the isothermal Jeans model does not capture the acceleration effect induced by tidal forces, as discussed previously.~\cite{Li:2025kpb} also found that, for halos of comparable mass, the projected enclosed mass decreases while the density slope becomes shallower after the turnover point. This trend differs from our collapsed GroupSIDM-147 halos, where the enclosed mass decreases but the slope steepens. A detailed comparison between our $N$-body cosmological simulations, the isothermal Jeans model predictions, and isolated simulations~\cite{Fischer:2025rky} is deferred to future work.

We also examine the case of field halos with substantial mass growth. Specifically, we select halos with the properties at $z=0.94$ comparable to the subhalos in the left panel of Fig.~\ref{fig:time-evo}, and track them to $z=0$, ensuring they never become subhalos of the main host. These halos experience numerous minor mergers, reaching $M_{\rm vir}=(8\textup{--}15)\times10^{10}~\rm M_\odot$ at $z=0$, corresponding to a growth factor of $4\textup{--}7$ relative to $z=0.94$. Their evolution, shown in the right panel of Fig.~\ref{fig:time-evo}, indicates that SIDM halos undergo continuous core expansion, characterized by decreasing enclosed mass and increasingly shallow density slopes approaching $\gamma_{\rm 2D}\sim0$. This behavior is consistent with expectations for halos of final mass $M_{\rm vir}\sim10^{11}~\rm M_\odot$, which should predominantly remain in the core-expansion phase under our SIDM models. These results suggest that sustained mass growth may delay core collapse, although a more detailed analysis is required to confirm this interpretation.

In this section, we have demonstrated the characteristic properties of SIDM halos relevant to strong lensing observables, using high-resolution cosmological simulations of a group system from the Concerto suite. These properties evolve dynamically in SIDM halos, in stark contrast to CDM halos, where they remain nearly static after formation. The stage of gravothermal evolution depends not only on the SIDM particle physics, through the self-interaction cross section, but also on the tidal environment and assembly history. These predictions can be tested with strong lensing observations across different redshifts. In particular, core-collapsed SIDM halos are expected to perturb gravitational lensing images more effectively than their CDM counterparts, as we discuss next.

\section{Perturber Analogs in Simulations}
\label{sec:subana}

We examine halos in our SIDM simulations at selected snapshots corresponding to four candidate strong gravitational lensing perturbers: {J0946}, {B1938}, {SDP.81}, and {SPT2147-50}. For comparison, we adopt perturber models from the latest studies as references.

\subsection{SDSS J0946\textup{+}1006}
\label{subsec: sdss}

\begin{figure*}[!]
    \centering
    \includegraphics[width=\linewidth]{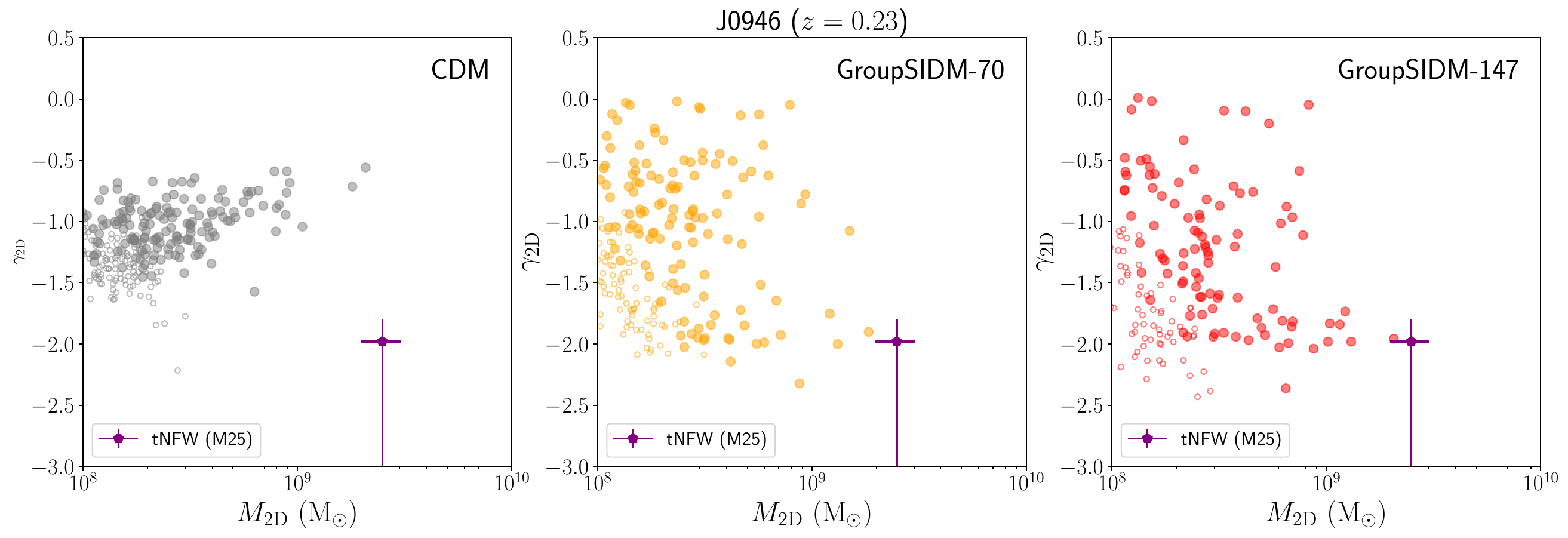}
    \caption{Projected enclosed masses and density slopes of subhalos ($R=1~{\rm kpc}$) at redshift $z=0.23$ in Group simulations for CDM (left, gray), GroupSIDM-70 (middle, orange), and GroupSIDM-147 (right, red). Subhalos resolved with more than $2000$ particles are shown as large filled markers, while those with $1000\textup{--}2000$ particles are shown as small open markers. For comparison, we show the inferred enclosed mass and density slope of the J0946 perturber from~\cite{Minor:2025} (purple), based on a truncated NFW fit including both multipole and kinematic priors; the error bar indicates the $95\%$ CL.}
    \label{fig:sub-j0946}
\end{figure*}

\begin{figure*}[!]
    \centering
    \includegraphics[width=\linewidth]{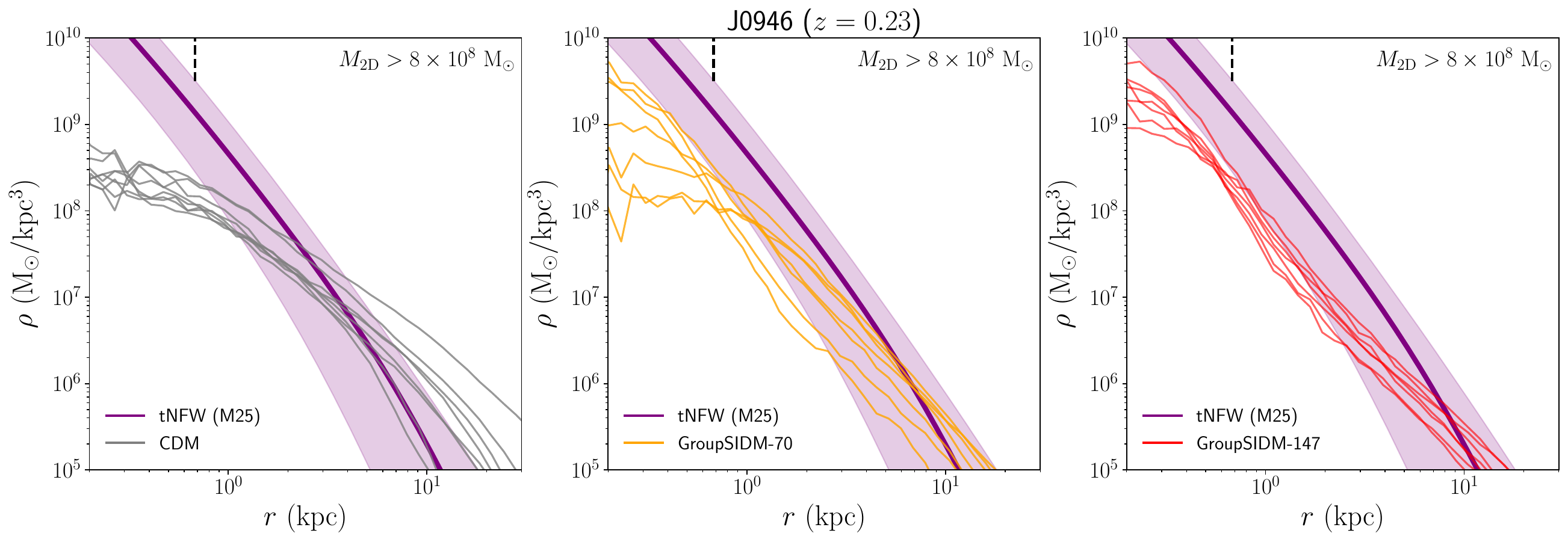}
    \caption{Density profiles of subhalos at redshift $z=0.23$ in Group simulations for CDM (gray), GroupSIDM-70 (orange), and GroupSIDM-147 (red), with projected masses $M_{\rm 2D}(1\,\mathrm{kpc})>8\times10^8~{\rm M_\odot}$. The vertical dashed line marks the resolution limit $2.8\epsilon = 0.67~\mathrm{kpc}$. For comparison, we show the reconstructed density profile of the J0946 perturber from~\cite{Minor:2025} (purple), based on a truncated NFW profile, with the shaded band denoting the $1\sigma$ uncertainty under Gaussian error assumptions for the marginalized parameters.}
    \label{fig:j0946-dens}
\end{figure*}

There have been multiple studies on possible perturber detection on J0946 in recent years~\cite{2010MNRAS.408.1969V, Minor:2020hic,Nightingale:2022bhh, Ballard:2023fgi, Enzi:2024ygw, Despali:2024ihn, Minor:2025, Tajalli:2025qjx}. The system is composed of a main foreground galaxy at $z=0.222$~\cite{Gavazzi:2008aq} and three background sources at $z=0.609$, $2.035$~\cite{Smith:2021oei}, and $5.975$~\cite{Collett:2020lii}. Although the exact perturber parameters differ slightly across these studies, they all concluded that the possible perturber is likely a subhalo of the main lens. Ref.~\cite{Nadler:2023nrd} showed that core-collapsed SIDM subhalos of the group host halo (Halo352) in the GroupSIDM-147 simulation can be significantly more dense than their CDM counterparts and that their properties are overall consistent with the perturber model in~\cite{Minor:2020hic}. 

In this work, we extend the previous study in several key aspects. First, we evaluate subhalo properties at the snapshot $z=0.23$, which is close to the redshift of the lens galaxy. By contrast, the analysis in Ref.~\cite{Nadler:2023nrd} was performed at $z=0$. The Concerto subhalo catalog we use is also more complete due to the addition of particle-tracking Symfind algorithm~\cite{Kong:2025kkt}. Second, we demonstrate that core-collapsed subhalos in the GroupSIDM-70 simulation can be sufficiently dense to serve as perturber candidates, even though the cross-section amplitude $\sigma_0/m$ is reduced by a factor of $2$. Lastly, we compare simulated subhalos with a more recent perturber model from Ref.~\cite{Minor:2025}, where two sources at different redshifts ({ $z=0.609$ and $z=2.035$}) are included in the lensing analysis; see also~\cite{Ballard:2023fgi}. In this case, the inferred enclosed mass within $1~{\rm kpc}$ is smaller than in the earlier work of Ref.~\cite{Minor:2020hic}, which considered only the lowest-redshift source. { We also note that constraints from Ref.~\cite{Ballard:2023fgi} are weaker than those in Ref.~\cite{Minor:2025}, with inferred concentrations at $\sim3\sigma$ and $\sim4\sigma$ above the cosmological median, respectively. As shown in Fig.~12 of Ref.~\cite{Minor:2025}, supersampling shifts the preferred concentration upward and reduces its uncertainty, likely explaining the difference between the two analyses. In the remainder of this discussion, we adopt the J0946 perturber model of Ref.~\cite{Minor:2025} as our fiducial benchmark.
}

Ref.~\cite{Minor:2025} used a truncated-NFW (tNFW) profile~\cite{Baltz:2007vq} to model the J0946 perturber
\begin{equation}
\rho(r)=\frac{\rho_{s}}{\left({r}/{r_{s}}\right)\left(1+{r}/{r_{s}}\right)^2\left(1+(r/{r_{{t}}})^2\right)},
\label{eq:tnfw}
\end{equation}
where $r_s$ and $\rho_s$ are the scale radius and density, respectively, and $r_t$ is the truncation radius. In the limit $r_t\rightarrow\infty$, this reduces to a regular NFW profile. For the lensing fit including the multipoles based on the $\alpha$-prior models, the inferred parameters are $\log(M_{200}/{\rm M_\odot})=9.84^{+1.80}_{-0.22}$ and $\log(c_{200})=2.64^{+0.70}_{-0.73}$, and $\log(r_t/{\rm kpc})=0.88^{+2.01}_{-1.57}$ ($95\%$CL)~\cite{Minor:2025}. Note here $M_{200}$ and $c_{200}$ are the mass and concentration of the subhalo would have if it were a field halo without tidal stripping, from which $r_s$ and $\rho_s$ in Eq.~\ref{eq:tnfw} can be reconstructed. The corresponding projected enclosed mass and density slope at $R=1~{\rm kpc}$ are $M_{\rm 2D}=2.49^{+0.54}_{-0.51}\times10^9~{\rm M_\odot}$ and $\gamma_{\rm 2D}=-1.98^{+0.18}_{-1.32}$ ($95\%$CL), respectively. 

In Fig.~\ref{fig:sub-j0946}, we present the projected enclosed masses and density slopes of simulated CDM and SIDM subhalos at redshift $z=0.23$, along with the inferred properties of the J0946 lensing perturber from~\cite{Minor:2025}. From the left panel, we see that CDM subhalos from the Group simulation cannot easily account for the steep slope inferred for the perturber. In contrast, core-collapsed SIDM subhalos in the GroupSIDM-70 and GroupSIDM-147 simulations shown in the middle and right panels, respectively, have a distribution of projected density slopes that are more consistent with the range inferred from the observations. Meanwhile, SIDM subhalos in the core-expansion phase have density slopes shallower than the CDM subhalos, as expected. More specifically, the $95\%$ ranges of the $\gamma_{\rm 2D}$ value are $(-1.41, -0.65)$, $(-2.01, -0.05)$, and $(-2.02, -0.01)$ for subhalos in the CDM, GroupSIDM-70, and and GroupSIDM-147 simulations, respectively. We find that GroupSIDM-70 and GroupSIDM-147 produce comparable diversity in subhalo density profiles, although low-mass subhalos in the latter tend to have slightly steeper slopes. Therefore, GroupSIDM-70 is also viable SIDM model to produce signatures in strong lensing observations.

Although the projected enclosed masses of core-collapsed subhalos are overall lower than that inferred for the observed perturber, a few of them have $M_{\rm 2D}>9\times10^9~\mathrm{M_\odot}$ and are compatible to the perturber model in~\cite{Minor:2025}, which includes two background sources at $z=0.609$ and $2.035$ in the lensing fit. If only the lowest-redshift source is included, the enclosed mass is higher by $34\%$ $M_{\rm 2D}\approx3.33\times10^{9}~{\rm M_\odot}$, while the density slope is shallower $\gamma_{\rm 2D}\approx-1.2$ \cite{Minor:2020hic}; see~\cite{Minor:2025} for a detailed discussion. Thus, our simulations are in better agreement with the latest perturber model. In this regard, an accurate lensing model is important for testing the SIDM scenario. 

On the simulation side, we note that the mass of the main halo (Halo352) is $1.3\times10^{13}~{\rm M_\odot}$, which lies at the lower end of the estimated mass range of the lens galaxy, $(1\textup{--}6)\times10^{13}~{\rm M_\odot}$~\cite{Minor:2020hic}. As a result, the abundance of simulated subhalos at the high-mass end could be somewhat underestimated. To test this, we analyzed another group host in the Concerto suite~\cite{Nadler:2025jwh}, Halo962, with a mass of $3.2\times10^{13}~{\rm M_\odot}$ for GroupSIDM-70. We find that the abundance of core-collapsed subhalos with $M_{\rm 2D}$ comparable to the perturber model is almost the same as in Halo352. In principle, Halo962 would accrete more massive subhalos than Halo352, but they may be in the core-expansion phase and subject to tidal disruption~\cite{Kong:2025kkt}. Other possible factors include the cosmological environments of the simulated group hosts, where higher-density regions tend to produce denser subhalos, as well as numerical issues associated with $N$-body simulations in the core-collapse regime, which may lead to an underestimation of the inner densities of collapsed subhalos~\cite{Zhong:2023yzk,Mace:2024uze,Palubski:2024ibb,Fischer:2024eaz,Fischer:2025rky}. Future work is needed to further address these points.

In Fig.~\ref{fig:j0946-dens}, we show density profiles of the subhalos at this snapshot ($z=0.23$) with a cutoff on the projected enclosed mass at $R=1~{\rm kpc}$ of $M_{\rm{ 2D}}> 8 \times 10^{8}~\rm{M_{\odot}}$, along with the density profile reconstructed from the truncated-NFW model~\cite{Minor:2025}. For the subhalos in our CDM simulation, their density profiles are too shallow in the inner regions $r\lesssim1~{\rm kpc}$, compared to the perturber model. In comparison, core-collapsed SIDM subhalos show better agreement with the perturber model, although their overall densities are slightly lower, with some cases falling within the model uncertainties. As the cross-section amplitude increases from $70~{\rm cm^2/g}$ to $147~{\rm cm^2/g}$, more subhalos evolve into a deeper collapse phase, resulting in higher densities in the inner regions. { We note that the inner density profiles of the core-collapsed subhalos lie at the lower end of the range favored by the perturber mass model, although their logarithmic density slopes match the inferred profile well. This is consistent with the somewhat low projected mass enclosed within $R=1~\mathrm{kpc}$ found in our simulations. The comparison suggests that the observed perturber may originate from a more massive and relatively high-concentration progenitor halo.} Additionally, lensing observables of the system $\gamma_{\rm 2D}$ and $M_{\rm 2D}$ evaluated at $R=1~{\rm kpc}$ are not sensitive to the mass distribution of the perturber at large radii ($r \gg 1~\rm{kpc}$); see also~\cite{Li:2025kpb}.

\begin{figure}[!]
    \centering
    \includegraphics[width=\linewidth]{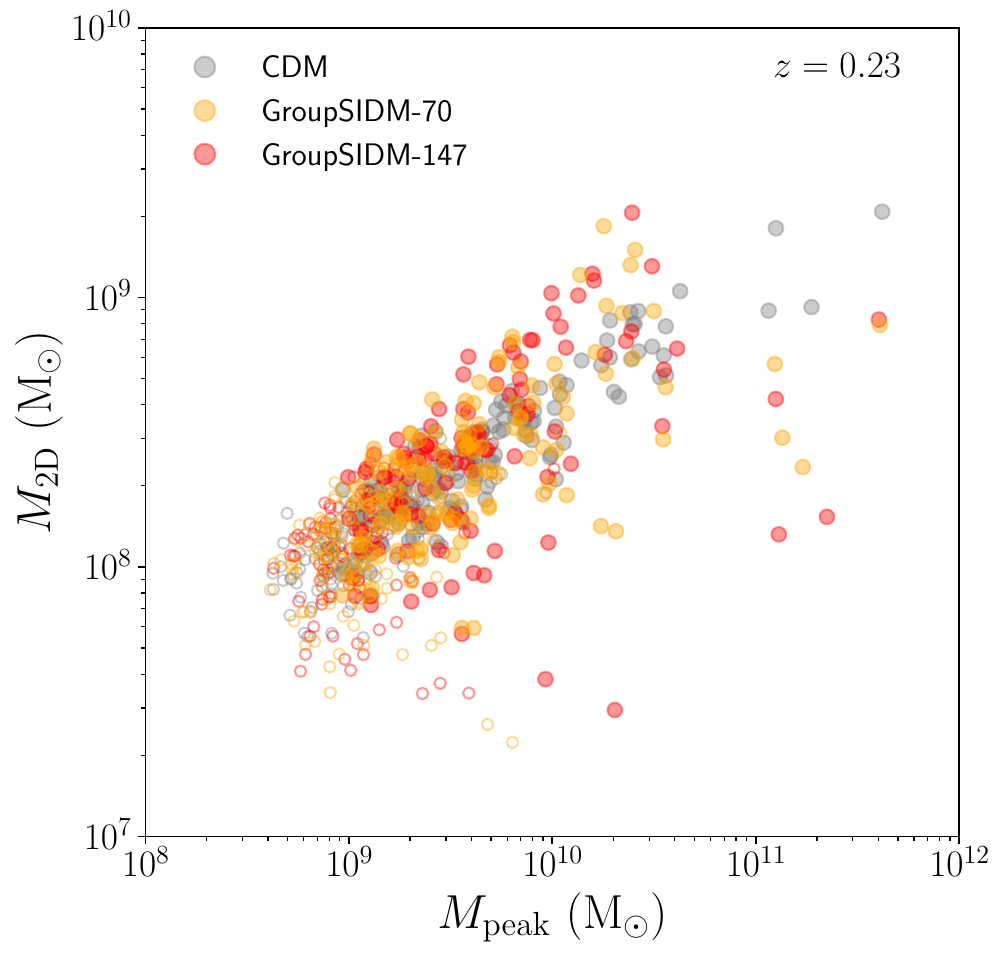}
    \caption{Projected enclosed mass $M_{\rm 2D}$ ($R=1~{\rm kpc}$) at redshift $z=0.23$ as a function of pre-infall peak mass of the simulated subhalos. Subhalos with more than $2000$ particles are shown with large filled markers, while those with $1000$ to $2000$ particles are shown with small open markers.}
    \label{fig:mpeak-z0d23}
\end{figure}

In Fig.~\ref{fig:mpeak-z0d23}, we show the projected enclosed mass as a function of the pre-infall peak mass of the subhalos. For the core-collapsed halos with $M_{\rm 2D}>10^9~{\rm M_\odot}$, the pre-infall peak masses lie in the range $(1\textup{--}3)\times10^{10}~{\rm M_\odot}$. We note that a few CDM subhalos also have $M_{\rm 2D}> 10^9~{\rm M_\odot}$, with peak masses exceeding $10^{11}~{\rm M_\odot}$; however, none of these subhalos are sufficiently dense in their inner regions to serve as perturber candidates. Using the peak masses shown in Fig.~\ref{fig:mpeak-z0d23}, we could apply abundance matching to infer the corresponding stellar masses of the subhalos and investigate their luminosities, thereby providing further constraints on the SIDM interpretation of the perturber. However, carrying out such an analysis would require a dedicated study of galaxy--halo connection in the SIDM framework, which is beyond the scope of this work. In what follows, we review recent progress in this direction and highlight some of the subtleties involved.

Assuming standard stellar mass--halo mass relations, Ref.~\cite{Li:2025kpb} argued that core-collapsed subhalos for the J0946 perturber would host luminous galaxies that should already have been detected. We note, however, that this constraint is subject to large uncertainties. Their semi-analytical analysis found that the required halo mass drops from $10^{11}~{\rm M_\odot}$ to $5\times10^{10}~{\rm M_\odot}$ as the halo concentration rises from the median to $3\sigma$ above it. In this case, the expected stellar mass decreases by nearly two orders of magnitude. As indicated in Fig.~\ref{fig:mpeak-z0d23}, our cosmological simulations show that the pre-infall masses of core-collapsed SIDM subhalos with $M_{\rm 2D}> 10^9~{\rm M_\odot}$ can be as low as $10^{10}~{\rm M_\odot}$, implying even smaller stellar masses. Moreover, core-collapsed SIDM subhalos necessarily pass through a core-expansion phase, which can enhance tidal mass loss of both dark matter and stars~\cite{Kong:2025kkt}. Consequently, the total stellar mass could be smaller than in the CDM case, although the stellar distribution within a core-collapsed subhalo may become more compact.

Ref.~\cite{He:2025wco} showed that an alternative lensing solution with light contamination from the perturber reduces the inferred concentration of the subhalo in the system; in this scenario, CDM subhalos may be dense enough to explain the data. A more robust, direct confirmation of the perturber's luminosity is needed, which can be achieved by analyzing the lensing system with multiple bands~\cite{He:2025wco}. If a luminous component is confirmed to be associated with this perturber, it would imply that a successful SIDM model cannot drive most or all subhalos into core collapse, but must instead produce a diversity of density profiles. Indeed, both the GroupSIDM-70 and GroupSIDM-147 simulations exhibit such a diverse distribution as shown in Fig.~\ref{fig:sub-j0946}. A dedicated study of SIDM simulations including baryons will further shed light on this situation.

\subsection{JVAS B1938\textup{+}666}
\label{subsec:jvas}

\begin{figure*}[!]
    \centering
    \includegraphics[width=\linewidth]{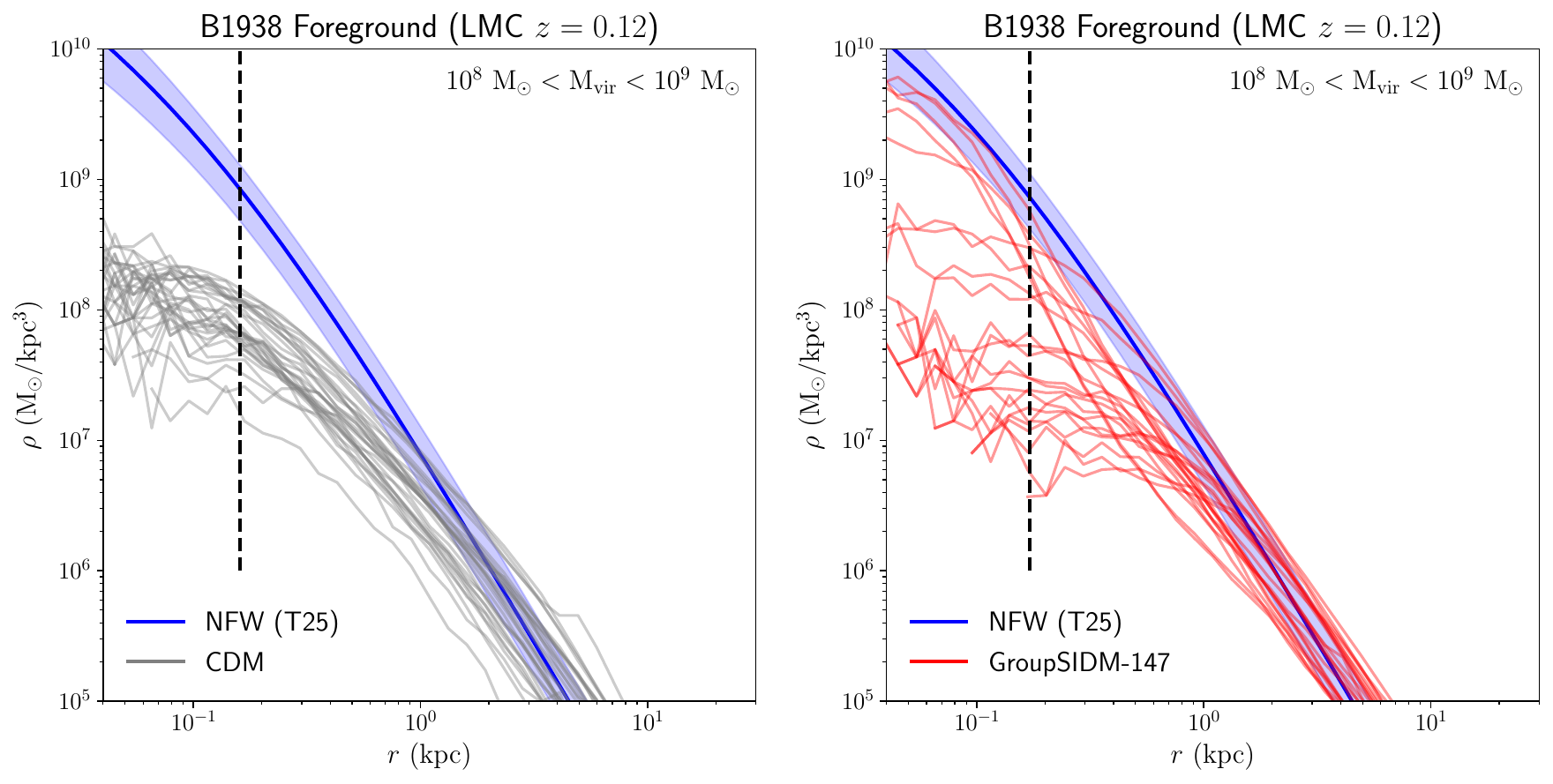}
    \caption{Density profiles of field halos at $z=0.12$ in LMC simulations for CDM (left, grey) and GroupSIDM-147 (right, red), with masses in the range $M_{\rm vir}=10^{8}$--$10^9~{\rm M_\odot}$. The perturber model (blue) assumes an NFW profile from~\cite{Tajalli:2025qjx}. The vertical dashed line marks the resolution limit $2.8\epsilon = 0.17~\rm{kpc}$.
    }
    \label{fig:b1938-front}
\end{figure*}

There have been multiple perturber models derived for the strong lensing system B1938, with recent studies respectively presenting a background perturber \cite{Sengul:2021lxe}, subhalo perturber~\cite{Despali:2024ihn}, and foreground perturber \cite{Tajalli:2025qjx}. For our study, we focus on the most recent foreground model~\cite{Tajalli:2025qjx} as the analysis is based on the latest and highest-resolution data available. Ref.~\cite{Tajalli:2025qjx} adopted an NFW profile and a power-law profile to model the foreground perturber and obtained constraints on their parameters by fitting to the lensing data. The total halo mass is well constrained for the NFW profile, but not for the power-law profile, and hence we choose the former as the reference to compare against our simulations. For the B1938 perturber, the halo mass is $M_{\rm vir}=5.08^{+3.86}_{-2.20}\times10^8~{\rm M_\odot}$ and the concentration is $c_{\rm vir}=185^{+116}_{-65}$ at $z\approx0.123$~\cite{Tajalli:2025qjx}. Taking the median values, the corresponding scale density and scale radius are $\rho_s\approx 7.74\times10^{9}~\rm{M_{\odot}~/kpc^3}$ and $r_s\approx 0.11~\rm{kpc}$, respectively.
    
The foreground perturber model from~\cite{Tajalli:2025qjx} reported a robust radius of $0.09~\rm{kpc}$, within which the perturber properties are confidently measured. This radius is much smaller than the spatial resolution of our Group simulations, which is set by the gravitational softening length as $2.8\epsilon = 0.67~\rm{kpc}$. Thus, we explore field halos in our LMC zoom-in simulation, which has a particle mass $m_{p}=6.3\times10^{3}~\rm{M_{\odot}}$ and a softening length of $\epsilon=0.06~\rm{kpc}$, providing a resolution limit of $2.8\epsilon=0.17~\rm{kpc}$; although this scale is still formally larger than the robust radius, it allows us to compare our simulated halos the inferred density profile much closer to the region constrained by lensing data.

As shown previously in Fig.~\ref{fig:ceff}, the properties of CDM field halos in the zoom-in region of the LMC simulation are consistent with those in the Group simulation. This indicates that the change of environment does not substantially bias our results. Furthermore, since the perturber in this model is a foreground halo at redshift $z=0.12$, which is significantly lower than the redshift of the lens galaxy ($z_{\rm lens}=0.881$), it is plausible that the perturber resides in a low-density region of the cosmic volume, consistent with the { cosmological} environment of the LMC simulations.

In Fig.~\ref{fig:b1938-front}, we present the density profiles of field halos in the LMC simulation for CDM and GroupSIDM-147, shown in the left and right panels, respectively. For comparison, we also include the inferred NFW density profile from~\cite{Tajalli:2025qjx}, along with its associated $1\sigma$ uncertainty bands (shaded regions). The vertical dashed line marks the resolution limit in our LMC simulations, $2.8\epsilon = 0.17~\rm{kpc}$. From the left panel, we note that the simulated CDM halos appear cored toward the center due to resolution limitations. Although their inner densities are systematically lower than the perturber model, we cannot conclude that CDM is in strong tension with the data. Ref.~\cite{Tajalli:2025qjx} extrapolated the inner density profiles of the simulated halos in the Illustris TNG50 CDM simulation using an analytic approach, and found analogs of the perturber, although they lie near the edge of the halo distribution in the relevant mass and redshift ranges. 

From the right panel, we see that the density profiles of the SIDM halos span a wider range. Some of them are deeply core-collapsed and consistent with the NFW model even at radii smaller than the resolution limit. We note that core-collapsed SIDM subhalos often have a substantial number of particles at radii $r<2.8\epsilon$, making their density profile measurements more reliable than CDM subhalos in this regime. Note that numerical issues associated with $N$-body simulations in the deep core-collapse phase, such as artificial heating, generally lead to underestimated inner densities~\cite{Zhong:2023yzk,Mace:2024uze,Palubski:2024ibb,Fischer:2024eaz,Fischer:2025rky}. In this sense, our results are likely conservative for core-collapsed systems.   

For the simulated halos shown in Fig.~\ref{fig:b1938-front}, their masses are in the range $M_{\rm vir}=10^8\textup{--}10^9~{\rm M_\odot}$. The SIDM halos closest to the perturber model have $M_{\rm vir} \sim 10^9~{\rm M_\odot}$; at this mass scale, the stellar mass is expected to be very low~\cite{DES:2019ltu,Benitez-Llambay:2020zbo}. Therefore, unlike J0946, which hosts a much more massive perturber, light contamination is likely less significant in the case of B1938. Our analysis suggests that core-collapsed SIDM halos are more likely candidates for the B1938 perturber than CDM halos, although the latter cannot be excluded given the systematic uncertainties of the density profile reconstruction and simulation predictions. Further work is needed to narrow down the favored model space.

A more recent study~\cite{Lei:2025pky} analyzed lensing data of the B1938 system under the assumption that the perturber is a subhalo of the lens galaxy, and found that the inferred density profile is actually cored, with a central density of $2.5\times10^7~{\rm M_\odot/kpc^3}$ and a core size of $0.5~{\rm kpc}$. Interestingly, some of our simulated SIDM halos, shown in the right panel of Fig.~\ref{fig:b1938-front}, align with this inferred cored profile. On the other hand, Ref.~\cite{Despali:2024ihn} assumed that the perturber is a subhalo and obtained a much steeper density profile. The discrepancy likely arises from their respective treatments of the subhalo density profile: Ref.~\cite{Despali:2024ihn} assumed a power-law density profile, whereas Ref.~\cite{Lei:2025pky} performed a non-parametric reconstruction.

\subsection{SDP.81}
\label{subsec:sdp}

\begin{figure*}[!]
    \centering
    \includegraphics[width=\linewidth]{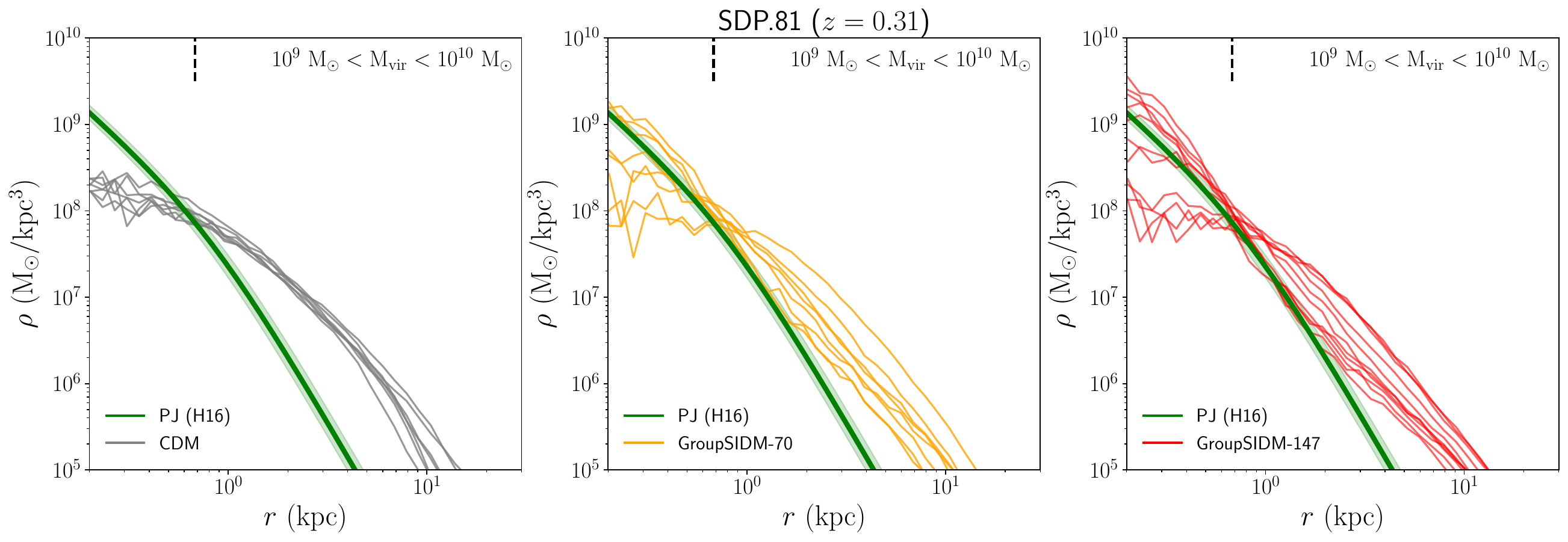}
    \caption{Density profiles of subhalos at $z=0.31$ in Group simulations for CDM (left), GroupSIDM-70 (middle), and GroupSIDM-147 (right), with masses in the range $10^{9}$--$10^{10}~{\rm M_\odot}$. The perturber model assumes a pseudo-Jaffe profile, reconstructed from~\cite{Hezaveh:2016ltk} (green). We note that a recent study~\cite{Stacey:2025} did not confirm the detection of a subhalo perturber in the SDP.81 system; our comparison is therefore provided as a case study and reference.}
    \label{fig:sdp81-dens}
\end{figure*}

Ref.~\cite{Hezaveh:2016ltk} reported the possible detection of a substructure with a mass of $M_{\rm sub} = 10^{8.96\pm0.12}~\rm{M_{\odot}}$ in the strong-lensing system SDP.81 using ALMA data. The lens is a massive elliptical galaxy at $z=0.2999$ and the source is a background star-forming galaxy at $z=3.042$~\cite{Wong:2015sfa}. A more recent study~\cite{Stacey:2025} re-analyzed data of the SDP.81 system and did not confirm evidence for a subhalo perturber. The earlier detection reported in~\cite{Hezaveh:2016ltk} may have been spurious, arising from inadequate modeling of multipoles in the annular lensing mass distribution. Interestingly, Ref.~\cite{Stacey:2025} also demonstrated that ALMA data are of sufficient quality to detect subhalos with masses below $10^{10}~{\rm M_\odot}$, if such subhalos are present. For this reason, we compare our simulated subhalos with the perturber model presented in~\cite{Hezaveh:2016ltk} as a case study. Should a detection of a subhalo with mass $10^9\textup{--}10^{10}~{\rm M_\odot}$ be confirmed in the future, our comparison will serve as a useful reference.  

Ref.~\cite{Hezaveh:2016ltk} used a pseudo-Jaffe profile \cite{1983MNRAS.202..995J, Munoz:2001bw} to model the perturber:
\begin{equation}
\rho(r)=\frac{\sigma^2_v}{2\pi G r^2_t }\frac{r_{\mathrm{t}}^4}{r^2\left(r^2+r_{{t}}^2\right)},
\label{eq:jaffe}
\end{equation}
where $\sigma_v$ is the 1D velocity dispersion of the subhalo, $r_{t}$ is the truncation radius, and $G$ is Newton's constant. From Eq.~\ref{eq:jaffe}, we can obtain the central projected surface density $\sigma^2_v/(2G)$ and the total mass $M=\pi\sigma^2_vr_t/G$. In order to reduce the number of free parameters, Ref.~\cite{Hezaveh:2016ltk} further assumed that $r_t=\left(\sigma_v / \sqrt{2} \sigma_G\right) r_{E}$, where $r_{E}$ is the Einstein radius of the main lens and $\sigma_{G}$ is its velocity dispersion. To reconstruct the density profile, we adopt $\theta_{E}=1.62~\rm{arcsec}$ from Table 2 in Ref.~\cite{2010Sci...330..800N}, equivalent to $r_{E}=7.06~\rm{kpc}$ following cosmology adopted by Ref.~\cite{2015PASJ...67...72T}, and $\sigma_G=265~{\rm km/s}$ from Table 1 in Ref.~\cite{2015PASJ...67...72T}, corresponding to their median values. The total inferred mass is $10^{8.96\pm0.12}~\rm{M_{\odot}}$~\cite{Hezaveh:2016ltk}. With these conditions, we determine as $\sigma_v\approx 39.4~\rm{km~s^{-1}}$ and $r_t\approx 0.8~\rm{kpc} $ in Eq.~\ref{eq:jaffe}.

In Fig.~\ref{fig:sdp81-dens}, we show density profiles of subhalos at $z=0.31$ from the Group CDM and SIDM simulations, together with the perturber model reconstructed from~\cite{Hezaveh:2016ltk}. For the simulated subhalos, their masses are within the range $10^{9}$--$10^{10}~\rm{M_{\odot}}$; the SIDM subhalos closest to the perturber's inferred density profile have masses of $2\times10^{9}~\rm{M_{\odot}}$. None of the CDM subhalos match the reconstructed perturber profile well. This is not surprising, as the perturber density profile scales as $r^{-2}$ towards the central regions, while the CDM subhalo inner profiles are expected to scale as $r^{-1}$, which is consistent with our results at scales larger than the resolution limit. In contrast, many of the core-collapsed SIDM subhalos in both GroupSIDM-70 and GroupSIDM-147 are dense enough to be candidates for the perturber. Although the simulated SIDM subhalos have higher densities than the assumed pseudo-Jaffe profile for $r>2~\rm{kpc}$, the actual profile of the perturber at these radii is likely not constrained by the current lensing data, as in the case J0946. A study that forward models the lensing signal using our simulated subhalo profiles would help clarify the radial sensitivity of the measurements, particularly if a detection is confirmed.

\subsection{SPT2147\textup{-}50}
\label{subsec:spt}

\begin{figure*}[!]
    \centering
    \includegraphics[width=\linewidth]{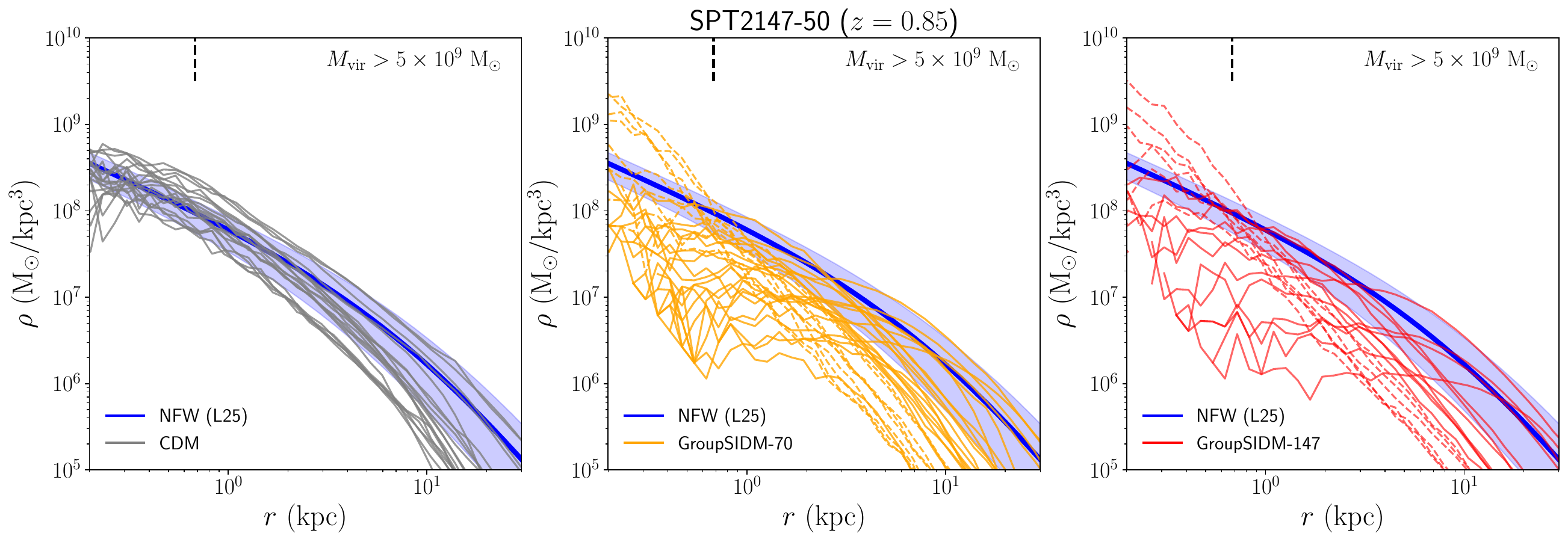}
    \caption{Density profiles of subhalos at redshift $z=0.85$ in Group simulations for CDM (left), GroupSIDM-70 (middle), and GroupSIDM-147 (right), with masses $M_{\rm vir}>5\times 10^9~{\rm M_\odot}$ (solid). The perturber model assumes an NFW profile with a median concentration from~\cite{Lange:2024pef} (blue). For comparison, we also include deeply core-collapsed subhalos with masses $\approx10^9~{\rm M_\odot}$ in the middle and right panels (dashed).}
    \label{fig:spt-dens}
\end{figure*}

Ref.~\cite{Lange:2024pef} reported the possible detection of a dark substructure in the strong lensing system SPT2147\textup{-}50 using JWST data. The lens galaxy lies at redshift $z = 0.845$, while the background source is at $z = 3.76$~\cite{Reuter:2020}. Their analysis showed that lensing models including a dark substructure are strongly favored, even after accounting for multipole perturbations of the main lens, corresponding to a $5\sigma$ detection. Assuming an NFW density profile with a median concentration, they inferred a substructure mass of $\log_{10}(M_{200}/{\rm M_\odot})=10.87^{+0.53}_{-0.71}$ from the F444W filter. It is important to emphasize, however, that the inferred mass depends sensitively on the assumed density profile and concentration~\cite{Despali:2017ksx,Despali:2024ihn}. In particular, adopting a more concentrated profile can lower the inferred substructure mass to $5\times10^{9}~{\rm M_\odot}$ or below~\cite{Lange:2024pef}.

In Fig.~\ref{fig:spt-dens}, we show density profiles of subhalos in the Group CDM and SIDM simulations, together with the NFW perturber model from~\cite{Lange:2024pef}. The simulated subhalos have masses $M_{\rm vir}>5\times10^9~{\rm M_\odot}$. We find multiple CDM subhalos with density profiles similar to the perturber model, as expected since the latter assumes a median concentration. By contrast, in both SIDM simulations, most subhalos with $M_{\rm vir}>5\times10^9~{\rm M_\odot}$ at $z=0.85$ ($6.5~{\rm Gyr}$ after the Big Bang) develop central cores, resulting in lower inner densities than their CDM counterparts. A few subhalos at the lower end of the mass range undergo core collapse and their central densities are comparable to the perturber model. Furthermore, for SIDM subhalos with masses around $10^9~{\rm M_\odot}$, some have already experienced deep core collapse by $z=0.85$, reaching central densities a few times higher than the perturber model at radii $\lesssim 1~{\rm kpc}$.

It is of great interest to re-analyze SPT2147\textup{-}50 using more compact density profiles for the substructure and to infer the corresponding mass. Evaluating the relative performance of different density profiles in fitting the lensing data can provide a further test of SIDM predictions. Among the four candidate detections of strong lensing perturbers, this system lies at the highest redshift, corresponding to roughly half the present age of the Universe. To explain core collapse at such an early epoch, the SIDM cross section must be sufficiently large on the relevant mass scale. In this regard, these analyses would further test our GroupSIDM-70 and GroupSIDM-147 models.


{ We have shown that core-collapsed SIDM halos attain higher central densities than their CDM counterparts, resulting in better agreement with the density profiles inferred for strong-lensing perturbers overall. To quantify this comparison, we compute $\chi^2$ values between the simulated subhalos and each perturber mass model.

For each perturber, we sample the inferred density profile, $O_i$, and its associated uncertainty, $\sigma_i$, at radii $r_i$. We adopt $10$ logarithmically spaced radii spanning the range from the halo center to a radius close to the reported robust radius, when available. For each simulated subhalo, we interpolate the density profile $\rho(r)$, using a cubic spline and evaluate

\begin{equation}
\chi^2=\sum_{i=1}^{10}\frac{\left[\rho(r_i)-O_i\right]^2}{\sigma_i^2}.
\end{equation}

For J0946, we use the radial range $0.1\textup{--}1~\mathrm{kpc}$. Among the subhalos shown in Fig.~\ref{fig:j0946-dens}, the best-fitting systems yield $\chi^2=30.2$ for CDM, $\chi^2=25.9$ for GroupSIDM-70, and $\chi^2=23.2$ for GroupSIDM-147. For B1938, we use the range $0.01\textup{--}0.9~\mathrm{kpc}$. Among the subhalos shown in Fig.~\ref{fig:b1938-front}, the minimum values are $\chi^2=22.3$ for CDM and $\chi^2=7.1$ for GroupSIDM-147.

For SDP.81, no robust radius has been reported, and we therefore adopt the range $0.1\textup{--}1~\mathrm{kpc}$. Among the subhalos shown in Fig.~\ref{fig:sdp81-dens}, the best-fitting halos have $\chi^2=115.3$ for CDM, $\chi^2=51.2$ for GroupSIDM-70, and $\chi^2=32.3$ for GroupSIDM-147. For SPT2147-50, we adopt the range $0.1\textup{--}1~\mathrm{kpc}$. Among the subhalos shown in Fig.~\ref{fig:spt-dens}, the minimum values are $\chi^2=6.5$ for CDM, $\chi^2=5.8$ for GroupSIDM-70, and $\chi^2=10.3$ for GroupSIDM-147. We caution that these calculations assume uncorrelated measurements. While differences in $\chi^2$ between CDM and SIDM models provide a useful measure of their relative performance, the absolute $\chi^2$ values should not be over-interpreted. 

Overall, the best-fitting SIDM subhalos generally provide a better match to the inferred perturber density profiles than their CDM counterparts, particularly for J0946, B1938, and SDP.81. These results quantitatively support the conclusion that gravothermally collapsed SIDM subhalos can more naturally reproduce the high central densities inferred for several strong-lensing perturbers. We note that the perturber model for SPT2147-50 is well matched by both the CDM and SIDM simulations. This is not unexpected, as Ref.~\cite{Lange:2024pef} adopted a CDM halo profile with a concentration close to the median CDM expectation. The primary goal of that work was to establish the presence of a dark substructure rather than to discriminate among different subhalo density profiles. More precise measurements of the internal structure of strong-lensing perturbers will provide increasingly powerful tests of the SIDM interpretation.

}

\section{Discussion and Conclusion}
\label{sec:discon}

In this work, we explored the lensing characteristics of halos from the zoom-in SIDM Concerto simulation suite~\cite{Nadler:2025jwh}. Our analysis spans CDM and two SIDM models with large, velocity-dependent dark matter self-interactions. We demonstrated that gravothermal evolution induced by self-interactions can increase projected enclosed masses and produce cuspy inner density slopes, providing a possible explanation for the high densities inferred from all currently known lensing perturber candidates detected via gravitational imaging. Across the perturber models we considered, the inferred $V_{\max}$ spans $15\textup{--}65~\rm{km/s}$; thus, these systems probe the velocity-dependent SIDM cross section in this velocity range. The subhalo properties predicted in the two SIDM models we considered are overall similar, suggesting that, for the range $\sigma_{0}/m=70\textup{--}147.1~\rm{cm^2/g}$, the turnover velocity is a critical factor that determines the evolution stage of massive subhalos. In particular, to explain the high density of the J0946 perturber, a $10^{10}~{\rm M_\odot}$ halo must be in the deep collapse phase, which sets a strong constraint on the turnover velocity $w\sim100~{\rm km/s}$.

In SIDM, the lensing characteristics $M_{\rm 2D}$ and $\gamma_{\rm 2D}$ evolve continuously following the gravothermal evolution of halos, in stark contrast with CDM, where these quantities are almost static aside from minor changes induced by tidal stripping. Thus, observations of strongly-lensed systems at different redshifts might be able to distinguish SIDM and CDM predictions. Furthermore, for core-collapsed SIDM halos, we found that the evolution of inner enclosed mass and density slopes can be described by a two-stage process that is connected by a turnover point that maximizes enclosed mass for a given halo. For subhalos, the evolution is further accelerated by tidal stripping and the turnover occurs earlier in time. This behavior may provide a maximum enclosed mass estimation, thus constraining the properties of halos that can act as efficient lensing perturbers.

We compared enclosed masses, density slopes, and density profiles of simulated halos against the perturber models from recent observational studies, which use analytical functions for modeling the perturber density profiles. Future work that forward models the lensing signal of halos from high-resolution and more realistic simulations in lensing environments is needed to confirm such analogs. In this context, we note that some of perturber models probe the enclosed mass at radii close to or smaller than the resolution limits of our simulations. For example, the robust radius of foreground models for system B1938 is around $0.09~\rm{kpc}$~\cite{Tajalli:2025qjx}, which is close to the spatial resolution limit of even our LMC simulations, and below that of the Group simulations. Although higher-resolution simulations will be needed to address this, we note that achieving such resolution in cosmological simulations of a strong-lensing environment is challenging in the near future. On the other hand, recent studies~\cite{Zhang:2024fib,Fischer:2025rky} suggest that using a King density profile model~\cite{1962AJ.....67..471K} can provide an accurate fit to the region that is below resolution for simulated halos. We leave an application of this technique to lensing perturber analogs for future work.

As discussed previously, the luminosity of lensing perturbers may also provide possible constraints on the search for SIDM perturber analogs, particularly for massive subhalos such as the J0946 perturber~\cite{Li:2025kpb,He:2025wco}. Several factors should be considered in this regard. If a luminous component is confirmed to be associated with this perturber, it implies that a successful SIDM model cannot have a very large cross section that makes most or all halos core-collapsed, but instead must allow diverse density profiles. On the other hand, if this perturber is indeed a dark subhalo, we may derive an upper bound for the possible subhalo mass by assuming the stellar--halo mass relation from hydrodynamic cosmological simulations, although the scatter may be large at these mass scales~\cite{Garrison-Kimmel:2016szj}.

{

The analysis presented in this work is based on dark-matter-only simulations. Energy injection from baryonic processes associated with galaxy formation in subhalo progenitors, as well as feedback from central supermassive black holes in lens galaxies, can alter the evolution of SIDM halos and may delay or even prevent gravothermal collapse. It is therefore important to assess the impact of these effects on the SIDM predictions. Recently, Ref.~\cite{Kong:2026piq} investigated the influence of baryonic feedback on the gravothermal evolution of SIDM halos using controlled simulations with an oscillating baryonic potential. They found that gravothermal collapse remains robust in high-concentration halos, even in the presence of strong feedback-induced energy injection. Such halos are the most plausible hosts of the strong-lensing perturbers considered here. These results suggest that the dense central regions produced by SIDM core collapse can survive realistic baryonic feedback, although dedicated hydrodynamical simulations of group-scale environments will be needed to establish the generality of this conclusion.}

We conclude that forthcoming observations of large samples of strong-lensing perturbers may provide a sensitive probe of velocity-dependent dark matter self-interactions. {Indeed, a $10^6~\mathrm{M_\odot}$ strong-lensing perturber has recently been discovered in the B1938 system~\cite{Powell:2025rmj,Vegetti:2026mmx}. Its inferred inner mass would correspond to a $\gtrsim10\sigma$ concentration outlier in CDM, whereas it is naturally reproduced by SIDM subhalos after undergoing deep collapse~\cite{Yu:2025tmp,Zhang:2026gur}. If similar dense perturbers are identified in future lensing surveys, they could provide compelling evidence for dark matter self-interactions.} Many such discoveries are expected from the \emph{Euclid} mission~\cite{ORiordan:2022qds}, complemented by the large samples of strong lenses anticipated from \emph{Rubin}~\cite{Shajib:2024yft} and \emph{Roman}~\cite{Wedig:2025idn}. A crucial aspect of this work will be a better understanding of strong lensing selection effects: if the observed perturbers only sample the most concentrated halos that contribute to the lensing signal, forward modeling is needed, since a comparison to the entire population of simulated halos at a given redshift may not be appropriate. Our simulation framework will be valuable, as it allows us to forward model the observed lensing signal, which we plan to pursue in future work. Thus, the combination of $N$-body simulations like SIDM Concerto with lensing data can serve as a valuable tool to test the self-interacting nature of dark matter. 

\bigskip

\begin{acknowledgments}

We thank participants of the Valencia2025 SIDM workshop for helpful discussion. HBY acknowledges support by the U.S. Department of Energy under grant No.~DE-SC0008541 and the John Templeton Foundation under grant No.~63599. The opinions expressed in this publication are those of the authors and do not necessarily reflect the views of the funding agencies.

\end{acknowledgments}

\bibliography{apssamp}

\appendix

\section{Halo Properties with an Alternative Concentration Definition}
\label{appex:cv}

\begin{figure}[!]
    \centering
    \includegraphics[width=\linewidth]{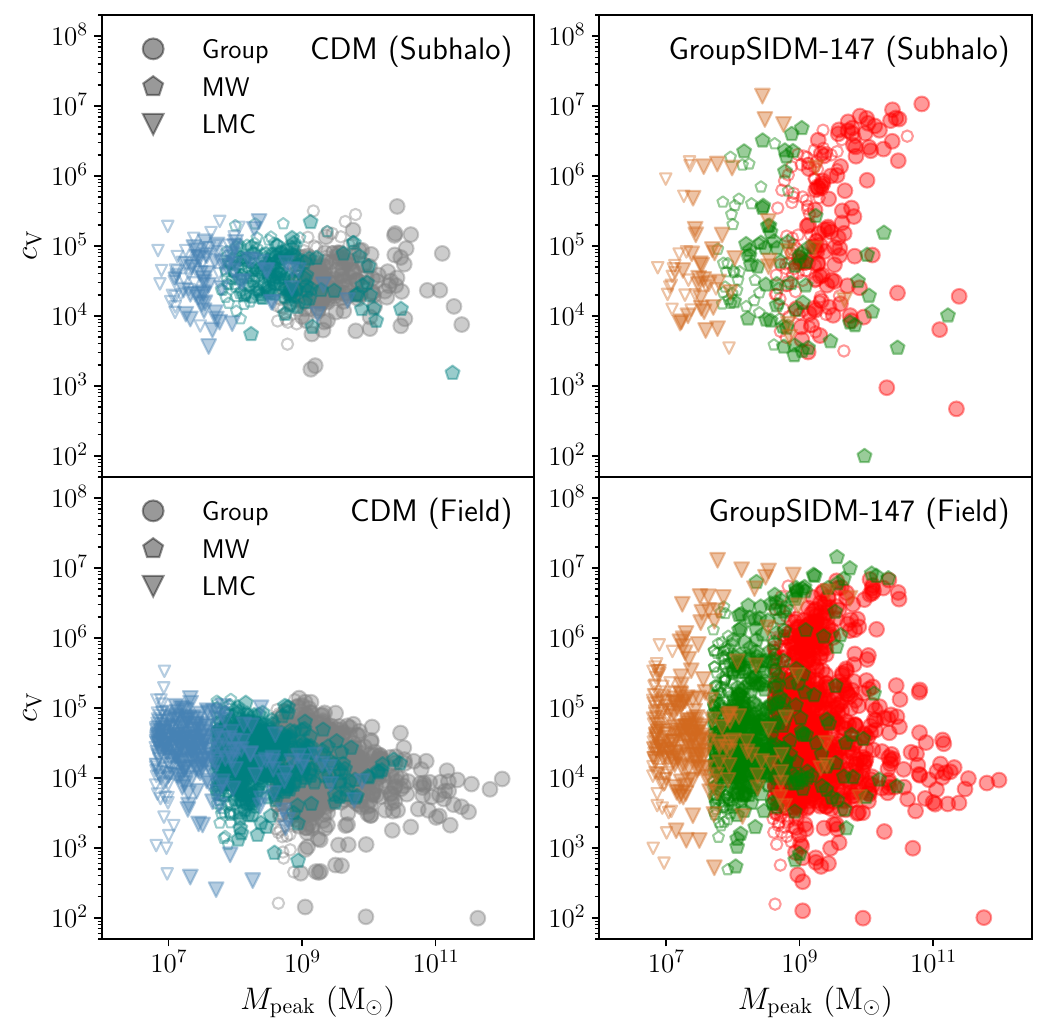}
    \caption{Distributions of the concentration (Eq.~\ref{eq:cv}) versus peak mass at $z=0$ for subhalos (top) and field halos (bottom) in the Group (circles), MW (pentagons), and LMC (triangles) simulations, shown for CDM (left) and GroupSIDM-147 (right). Halos with more than $2000$ particles are shown as large filled markers, and those with $1000\textup{--}2000$ particles as small open markers.}
    \label{fig:cv}
\end{figure}

\begin{figure*}[!]
    \centering
    \includegraphics[width=\linewidth]{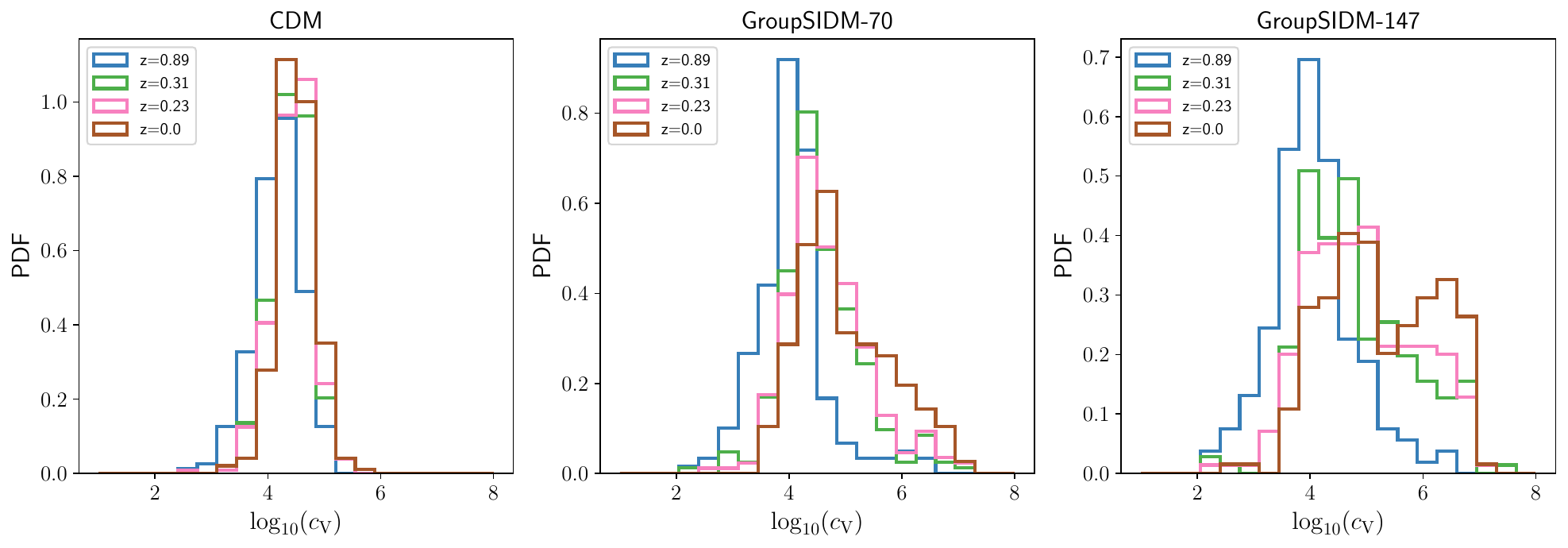}
    \caption{Probability distributions of the concentration (Eq.~\ref{eq:cv}) at redshifts $z = 0.89$ (blue),
$0.31$ (green), $0.23$ (pink), and $0$ (brown) in our Group zoom-in simulations for CDM (left), GroupSIDM-70 (middle), and
GroupSIDM-147 (right). We include all subhalos with more than $1000$ particles at each snapshot, equivalent to a subhalo mass larger than $4\times10^8~{\rm M_\odot}$.}
    \label{fig:cvevo}
\end{figure*}

\begin{figure*}[!]
    \centering
    \includegraphics[width=\linewidth]{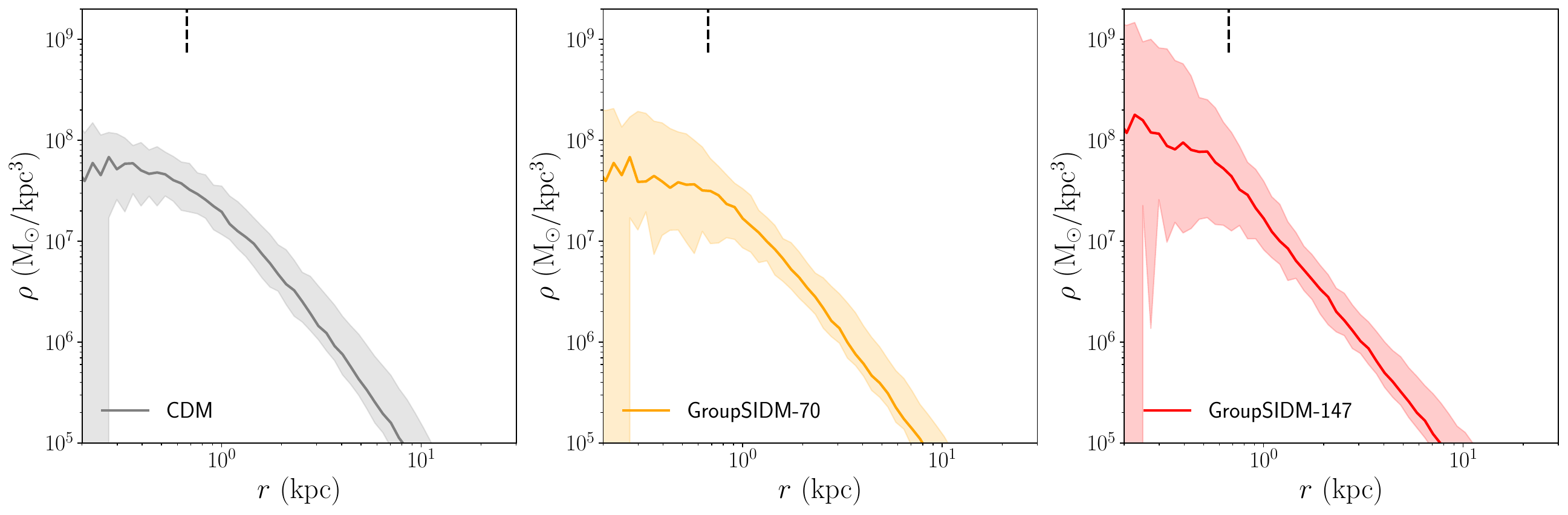}
    \caption{{ Median 3D density profiles (solid curves) with $1\sigma$ scatter (shaded regions) for subhalos of mass $10^{9}\textup{--}10^{10}~\mathrm{M_\odot}$ at $z=0$ in the Group CDM (left), GroupSIDM-70 (middle), and GroupSIDM-147 (right) simulations. The vertical dashed line marks the resolution limit $2.8\epsilon=0.67~\mathrm{kpc}$.} }
    \label{fig:medprofile}
\end{figure*}

An alternative and widely used definition of halo concentration is~\cite{Springel:2008cc}
\begin{equation}
c_V = 2\left(\frac{V_{\rm max}}{H(z)r_{\rm max}}\right)^2,
\label{eq:cv}
\end{equation}
where $H(z)$ is the Hubble parameter, $V_{\rm max}$ is the maximum circular velocity, and $r_{\rm max}$ is its corresponding radius. This definition is particularly useful for subhalos, since it does not rely on virial quantities. In what follows, we present the distribution and redshift evolution of $c_V$ for our simulated CDM and SIDM halos.

Fig.~\ref{fig:cv}, analogous to Fig.~\ref{fig:ceff}, shows the distributions of $c_V$ for subhalos and field halos. The trends mirror those in Fig.~\ref{fig:ceff}: SIDM halos display much larger scatter in $c_V$ at fixed $M_{\rm peak}$ than CDM, with core-collapsed halos reaching very high $c_V$ values. The scatter is comparable across all four host masses, suggesting little environmental bias. Notably, Ref.~\cite{Tajalli:2025qjx} estimated $\log c_V \sim 7$ for the J0946 perturber, which is well above the values reached by our simulated CDM subhalos, but consistent with SIDM subhalos in the deeply collapsed phase, as shown in the top panels of Fig.~\ref{fig:cv}.

Fig.~\ref{fig:cvevo} shows the probability distributions of concentration (Eq.~\ref{eq:cv}) at redshifts $z=0.89$ (blue), $0.31$ (green), $0.23$ (pink), and $0$ (brown) in the Group zoom-in simulations for CDM (left), GroupSIDM-70 (middle), and GroupSIDM-147 (right). We include all subhalos masses larger than $4\times10^8~{\rm M_\odot}$ at each snapshot. For CDM, the $c_V$ distributions are relatively narrow and evolve little with time. In contrast, SIDM subhalos exhibit broader distributions that shift toward higher $c_V$ values at lower redshifts, reflecting the increasing fraction undergoing core collapse. The trend strengthens with cross-section amplitude: in GroupSIDM-147, the $c_V$ distribution at $z=0$ is bimodal, with a substantial population of deeply collapsed subhalos at $\log(c_V)\gtrsim6$. This behavior closely parallels the evolution of $\gamma_{\rm 2D}$ in Fig.~\ref{fig:time-evo}.

{
\section{Subhalo Density Profiles}
\label{appex: medrho}

Besides the representative examples discussed in Sec.~\ref{subsec: rep}, we compute the median 3D density profiles and the corresponding $1\sigma$ scatter at $z=0$ for all subhalos with masses in the range $10^{9}\textup{--}~10^{10}~\mathrm{M_\odot}$ in our Group simulations, as shown in Fig.~\ref{fig:medprofile}. Relative to CDM, SIDM subhalos exhibit a substantially larger spread in their inner density profiles due to gravothermal evolution, which drives both core expansion and core collapse. These processes amplify the intrinsic halo-to-halo scatter present in CDM, increasing the $1\sigma$ dispersion in the central regions relevant for strong-lensing constraints by nearly an order of magnitude. Interestingly, the median central density in GroupSIDM-70 is comparable to that in CDM because the subhalo population contains both core-expanded and core-collapsed systems, whose contributions partially offset each other in the median. By contrast, the median central density in GroupSIDM-147 is higher, as most subhalos have progressed to a more advanced stage of gravothermal collapse.

}

\end{document}